\documentclass[]{spie}  %>>> use for US letter paper
\usepackage{hyperref}
\usepackage{graphicx}
\usepackage{amsmath,amsfonts,amssymb}

\title{The Cosmology Large Angular Scale Surveyor\\Receiver Design}

\author[a]{Jeffrey~Iuliano}
\author[a]{Joseph~Eimer}
\author[a,b]{Lucas~Parker}

\author[a]{Gary~Rhoades}

\author[a,c]{Aamir~Ali}
\author[a]{John~W.~Appel}
\author[a]{Charles~Bennett}
\author[a]{Michael~Brewer}
\author[e]{Ricardo~Bustos}
\author[f]{David~Chuss}
\author[a]{Joseph~Cleary}
\author[a]{Jullianna~Couto}
\author[a]{Sumit~Dahal}
\author[d]{Kevin~Denis}
\author[g]{Rolando~D\"{u}nner}

\author[d]{Thomas~Essinger-Hileman}
\author[g]{Pedro~Fluxa}
\author[i]{Mark~Halpern}
\author[a]{Kathleen~Harrington}
\author[d]{Kyle~Helson}
\author[h]{Gene~Hilton}
\author[i]{Gary~Hinshaw}
\author[h]{Johannes~Hubmayr}
\author[a]{John~Karakla}
\author[a]{Tobias~Marriage}
\author[a]{Nathan~Miller}
\author[k]{Jeffrey~John~McMahon}
\author[a]{Carolina~Nu\~{n}ez}
\author[a]{Ivan~Padilla}
\author[j]{Gonzalo~Palma}

\author[a]{Matthew~Petroff}
\author[j]{Bastian~Pradenas~M\'{a}rquez}
\author[l]{Rodrigo~Reeves}
\author[h]{Carl~Reintsema}
\author[d]{Karwan~Rostem}
\author[a]{Deniz~Augusto~Nunes~Valle}
\author[a]{Trevor~Van~Engelhoven}
\author[a]{Bingjie~Wang}
\author[a]{Qinan~Wang}
\author[a]{Duncan~Watts}
\author[a]{Jenet~Weiland}
\author[d]{Edward~J.~Wollack}
\author[a,m]{Zhilei~Xu}
\author[i]{Ziang~Yan}
\author[n]{Lingzhen~Zeng}

\affil[a]{Department of Physics and Astronomy, Johns Hopkins University, Baltimore, MD 21218, USA}

\affil[b]{Space and Remote Sensing, MS D436, Los Alamos National Laboratory, Los Alamos, NM 87544, USA}

\affil[c]{Department of Physics, University of California, Berkeley, CA 94720, USA }
\affil[d]{NASA Goddard Space Flight Center, Greenbelt, MD 20771, USA }
\affil[e]{Facultad de Ingenier\'{i}a, Universidad Cat\'{o}lica de la Sant\'{i}sima Concepci\'{o}n, Alonso de Ribera 2850, Concepci\'{o}n, Chile }
\affil[f]{Department of Physics, Villanova University, Villanova, PA 19085, USA }
\affil[g]{
Instituto de Astrof\'{i}sica and Centro de Astro-Ingenier\'{i}a, Facultad de F\'{i}sica, Pontificia Universidad Cat\'{o}lica de Chile, 7820436
Macul, Santiago, Chile }
\affil[h]{National Institute of Standards and Technology, Boulder, CO 80305, USA}
\affil[i]{Department of Physics and Astronomy, University of British Columbia, Vancouver, BC V6T 1z4 ,Canada}
\affil[j]{Departmento de F\'{i}sica, FCFM, Universidad de Chile, Blanco Encalada 2008, Santiago, Chile}

\affil[k]{Department of Physics, University of Michigan, Ann Arbor, MI, 48109, USA}
\affil[l]{Departamento de Astronom\'{i}a, Universidad de Concepci\'{o}n, Casilla 160 C, Concepci\'{o}n, Chile}
\affil[m]{Department of Physics and Astronomy, University of Pennsylvania, Philadelphia, PA 19104, USA}
\affil[n]{Harvard-Smithsonian Center for Astrophysics, Cambridge, MA 02138, USA }

\authorinfo{Further author information: Send correspondence to Jeffrey Iuliano (jiulian1@jhu.edu)}

\usepackage{ads_abbrevs}

\newcommand{\ssection}[1]{\section{#1}}
\newcommand{\ssubsection}[1]{\subsection{#1}}

\usepackage[title]{appendix}
\usepackage{url}
\usepackage{siunitx}
\usepackage{multirow}
\usepackage{multicol}

\begin{document} 
\maketitle

\begin{abstract}
The Cosmology Large Angular Scale Surveyor consists of four instruments performing a CMB polarization survey.  Currently, the \SI{40}{GHz} and first \SI{90}{GHz} instruments are deployed and observing, with the second \SI{90}{GHz} and a multichroic 150/\SI{220}{GHz} instrument to follow.  The receiver is a central component of each instrument's design and functionality.  This paper describes the CLASS receiver design, using the first \SI{90}{GHz} receiver as a primary reference.  Cryogenic cooling and filters maintain a cold, low-noise environment for the detectors.  We have achieved receiver detector temperatures below \SI{50}{mK} in the \SI{40}{GHz} instrument for 85\% of the initial 1.5 years of operation, and observed in-band efficiency that is consistent with pre-deployment estimates.  At \SI{90}{GHz}, less than 26\% of in-band power is lost to the filters and lenses in the receiver, allowing for high optical efficiency.  We discuss the mounting scheme for the filters and lenses, the alignment of the cold optics and detectors, stray light control, and magnetic shielding.

\end{abstract}


\ssection{Introduction}
	The Cosmic Microwave Background (CMB) is a rich source of information about the universe.  The temperature anisotropy has been well mapped and understood, revealing details of the early universe and constraining cosmological models and parameters\cite{WMAP,Planck}.  CMB polarization is a similarly useful tool, which requires further work to fully explore\cite{TCR}.

	The CMB was emitted during the epoch of last scattering.  Before this time, the radiation in the universe was tightly coupled with the matter.  As the matter formed neutral hydrogen, it rapidly decoupled from the radiation, which became the CMB.  Quadrupole anisotropies around a point on this surface of last-scattering produce a net CMB polarization through Thomson scattering.  These quadrupole anisotropies can be sourced from density (scalar) fluctuations in the matter-energy material of the early universe or a primordial gravitational-wave (tensor) background, which is generically predicted by inflationary theories\cite{Guth,KKS}.  After reionization, when the hydrogen in the universe became re-ionized (likely due to light from the first stars), CMB photons scatter off newly free electrons, imprinting a further polarization pattern with information about this period in the universe's history.  The CMB polarization pattern can be decomposed into a basis of E and B modes, which are analogous to patterns of electric and magnetic fields.  While the gravitational wave background produces both E and B modes, the scalar perturbations can only produce E-mode polarization\cite{SZ}.  So, B-mode CMB polarization is evidence of the primordial gravitational wave background and can be used to characterize inflation\cite{Kamionkowski}.

	The Cosmology Large Angular Scale Surveyor (CLASS) is located on Cerro Toco inside Parque Astronómico de Atacama in Chile.  CLASS is carrying out a CMB polarization survey optimized for large angular scales, where the CMB polarization anisotropy is largest and has interesting features (the reionization and recombination peaks)\cite{katie,tom,eimer_optics}.  One goal of CLASS is to detect and characterize the primordial B-mode power spectrum from inflationary gravitational waves, with sensitivity to $r$ (the tensor-to-scalar ratio) as low as 0.01\cite{Watts}.  CLASS will also perform a nearly cosmic-variance limited measurement of the E-mode power spectrum.  This will give a sample-variance limited estimate of $\tau$, the optical depth to reionization, which will determine when the first stars formed and help constrain neutrino masses\cite{Duncan_tau}.

% \ssection{Survey Strategy and Instrument Overview}
	A CLASS-like survey has three primary challenges.  First, the CMB polarization anisotropy is small, so CLASS needs high sensitivity.  Ref.~\citenum{Sumit} discusses the detector design and current progress: the detectors and receiver are optimized for high optical efficiency to reduce integration time needed to achieve the desired sensitivity\cite{NASA1,NASA2,John_SPIE}.  The second challenge is recovering large angular scales from the ground, where $1/f$ noise from atmospheric and thermal drifts can dominate the CMB signal.  To address this, CLASS developed a front-end polarization modulator -- the Variable-delay Polarization Modulator (VPM) -- which is the first element in the CLASS optical system.  Ref.~\citenum{katie} discusses the VPM design, function, and current status.  By modulating the polarization signal, CLASS is able to reliably recover large angular scales\cite{Harrington2018,VPM1,VPM2,Miller}.  Finally, CLASS will have to remove the effects of polarized galactic foregrounds to recover the CMB polarization signal\cite{Watts,Duncan_tau}.  To help remove these foregrounds, CLASS will observe across multiple frequency bands (\SI{40}{GHz}, \SI{90}{GHz}, \SI{150}{GHz}, and \SI{220}{GHz}).  Currently, the \SI{40}{GHz} and first \SI{90}{GHz} instruments are deployed, with the second \SI{90}{GHz} instrument and the multichroic 150/\SI{220}{GHz} instrument following soon.

	After the VPM, incoming light reflects off two mirrors, and enters the receiver.  The receiver provides a cold and stable environment for the detectors -- this is critical to achieving low-noise, high-sensitivity observations.  The receiver is also an optical system designed to maintain the alignment of the lenses and detectors, and to control stray light.  In this paper, we discuss the cryogenic performance of the CLASS receivers, and how that performance is maintained through a robust (but high efficiency) filtering scheme.  We also describe the mounting and alignment of the optical elements, stray-light controls, and magnetic shielding.  Finally, with the first instruments in the field, we report on the cryogenic performance of the deployed instruments while observing.  We focus primarily on the \SI{90}{GHz} receiver design, but note the key differences between the \SI{90}{GHz} and \SI{40}{GHz} instruments and expected alterations for the High-Frequency (HF, 150/\SI{220}{GHz}) system.

\ssection{Design Requirements and Overview}
	The CMB is a \SI{2.725}{K} blackbody\cite{COBE} with hundreds-of-milliKelvin-level temperature anisotropy.  The E-mode polarization pattern is anisotropic at a microKelvin level, while the primordial B-mode pattern corresponds to anisotropy at \SI{e-2}{\mu K} or smaller.  Thus, CLASS depends on high survey sensitivity, which the receiver design supports.  In particular, the receiver maintains a cold environment for the detectors.  The Transition Edge Sensor (TES) bolometers operate with a superconducting transition at $\sim$1\SI{50}{mK}, but the receiver needs to provide a thermal bath for the detectors below $\sim$\SI{70}{mK} to keep detector saturation power above optical loading (and help limit thermal noise)\cite{Sumit}.

	In factory configuration, the CLASS dilution refrigerators can maintain the mixing chamber (which sets the detector bath temperature) at \SI{25}{mK}.  In its final configuration, however, the receiver is open to the sky, introducing additional optical loading, while wiring and optical supports introduce additional conductive loading -- both of which increase the cold stage temperature.  The system can accommodate $\sim$\SI{150}{\mu W} of additional power while keeping the cold stage below \SI{70}{mK} (see Figure \ref{cryopower}).  Given that $\sim$\SI{70}{W} of optical power enters through the receiver's window (the vast majority of it out-of-band), a robust filtering scheme is necessary.  The filters reduce the power reaching the cold stage by a factor of more than $10^5$, but have minimal impact in-band.  Similarly, mechanical supports and readout wiring do not add significant conductive loading.

	The receiver is also an optical system, including lenses (cold refractive optics are desirable because they emit less in-band) and detectors.  Each receiver houses three main optical elements -- the \SI{4}{K} lens, the \SI{1}{K} lens, and the detectors (assembled into the focal plane).  The lenses and filters are built into optical assemblies (or \emph{stacks}), which are mounted at the end of each of the radiation shields. The receiver is responsible for the alignment of three primary optical components (the two lenses and the focal plane), which are also the components most sensitive to mis-alignment.  It is also important to control stray light and reflections inside the receiver.

	\begin{figure*}[h]%
		\begin{center}
		\includegraphics[width=.8\columnwidth]{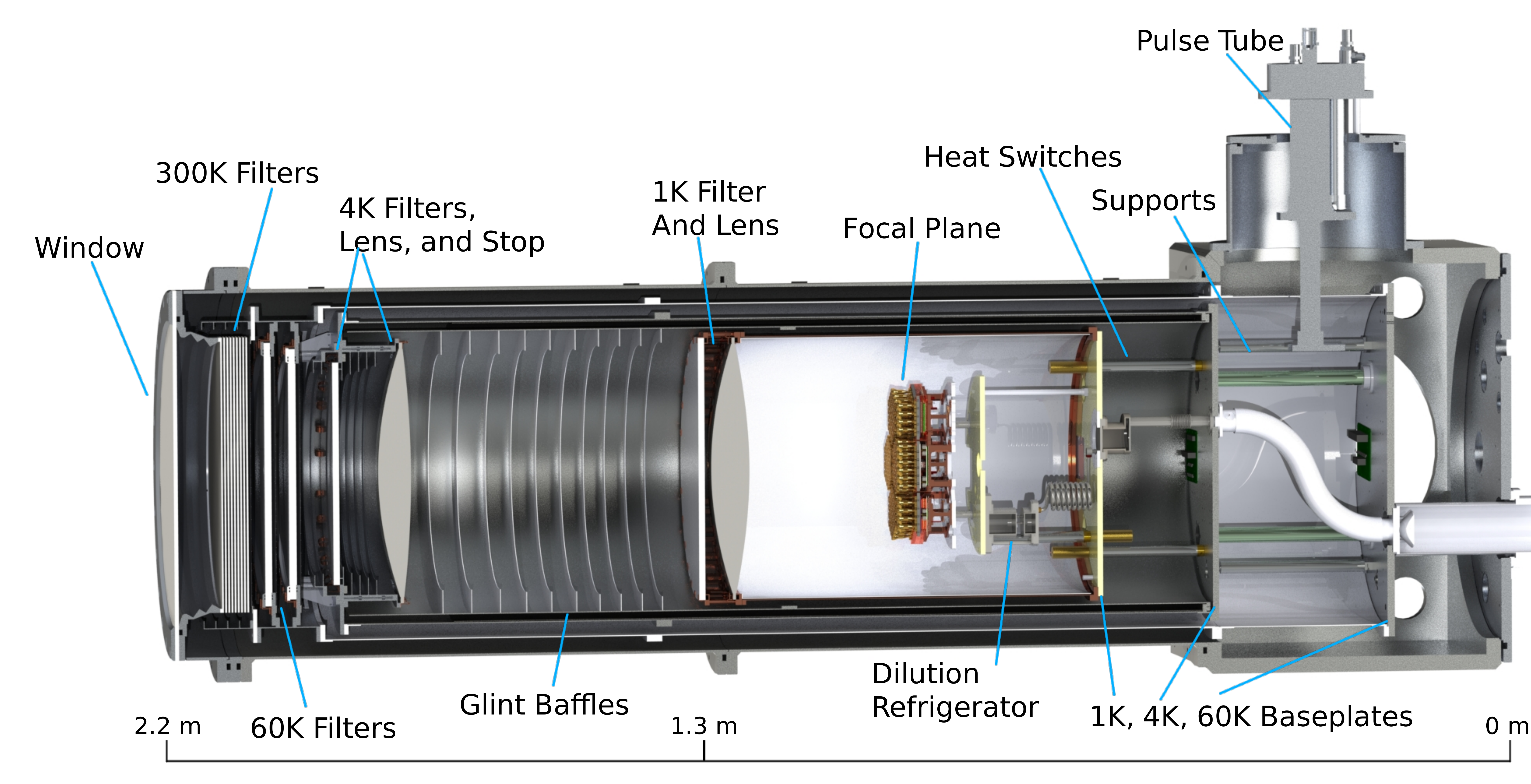}%
		% \\HF DIAGRAM?
		\end{center}
		\caption{Diagram of the \SI{90}{GHz} receiver design. See Ref.~\citenum{katie} for a similar diagram of the \SI{40}{GHz} receiver.  }%Diagrams of the \SI{40}{GHz} (top) and \SI{90}{GHz} (bottom) receiver designs. 
		\label{diagrams}%
		\vspace{2mm}
	\end{figure*}

	There are four CLASS receivers: \SI{40}{GHz} (Q-band), \SI{90}{GHz} (W-band 1), \SI{90}{GHz} (W-band 2), and 150/\SI{220}{GHz} (High-Frequency).  While the different frequencies (and optical designs) require some changes between the receivers, the overall design of the four receivers is largely similar.  Figure \ref{diagrams} shows the design of the \SI{90}{GHz} receivers.  Almost every component of the \SI{40}{GHz} receiver is qualitatively identical to the \SI{90}{GHz} design, with slightly different sizes and shapes to account for the different frequency and optical design.  The primary difference is the addition of piano wire supports (see Section \ref{PianoWireSupports}) for the \SI{90}{GHz} instrument to aid in maintaining optical alignment (due to tighter tolerances).  The High-Frequency receiver design is essentially the same as the \SI{90}{GHz} design, but employs silicon rather than plastic lenses.  The filter designs between receivers differ slightly, reflecting both developments in understanding and frequency-dependent filter characteristics.

\ssection{Cryogenic System Overview}

	The base on which the CLASS receivers are built is a custom-made BlueFors\cite{bf} cryostat, which can be seen in Figure \ref{diagrams}.  The cryostat has two cooling systems -- a Pulse Tube (PT) and a Dilution Refrigerator (DR) -- and is separated into five distinct temperature stages: \SI{300}{K}, \SI{60}{K}, \SI{4}{K}, \SI{1}{K}, and \SI{100}{mK}.  The \SI{300}{K} stage forms the superstructure of the system and is the vacuum vessel.  Each of the other stages consists primarily of a baseplate, to which the relevant cooling mechanism is attached, and a cylindrical radiation shield (or shell) that launches off the baseplate -- except the \SI{100}{mK} baseplate, to which only the detectors are attached.  See Table \ref{stages} for operating temperatures and cooling capacities.

	The \SI{60}{K} and \SI{4}{K} shells are wrapped in layers (30 and 10 respectively) of aluminized mylar superinsulation -- a type of Multi-Layer Insulation (MLI).  These sheets cut the power reaching the shell by approximately $1/N$, where $N$ is the number of layers\cite{mli}.  It is difficult and time-consuming to wrap each layer individually (which was the approach for the \SI{40}{GHz} receiver).  So, for the \SI{90}{GHz} and HF receivers, we used MLI that comes bonded in ten-layer sheets\cite{tape}.  The ten-layer sheets are connected intermittently by point fuses between the layers, making them significantly easier to work with at the cost of a small increase in conduction between layers.  The MLI is held in place with reflective tape (Coolcat B-R50 \cite{tape}) and arranged so that the seams of each ten-layer sheet are non-overlapping (and do not interfere with bolts on the ends of the shells).  The warm side of the \SI{60}{K} baseplate also has 30 layers of MLI, with slits cutout to avoid supports and wires.  Each of the ten-layer sheets has differently arranged slits to minimize overlap, and is applied by temporarily attaching the sheet to a cardstock template.  Proper arrangement and taping of the insulation are important as even a small light leak can add significant loading and prevent proper cooling of the system.

	\begin{table}[ht]%
	\begin{center}
		\begin{tabular}{c|llll}
		\multirow{2}{*}{Stage} 	& Factory 			&Deployment		& Cooling  & \multirow{2}{*}{Cooler}\\
								& Configuration  	&Configuration	& Capacity \\
													
		\hline
		\SI{300}{K} 	&	\SI{280}{K} 		&\SI{280}{K} 		& \\
		\SI{60}{K} 	&	\SI{36}{K} 		&\SI{40}{K} 		&	\SI{40}{W} 	& PT\\
		\SI{4}{K} 	& 	\SI{2.8}{K} 		&\SI{3}{K} 		& 	\SI{1.5}{W} 	& PT\\
		\SI{1}{K} 	& 	500-\si{800}{mK} 	&500-\SI{800}{mK} & 	\SI{100}{mW} 	& DR\\
		\SI{100}{mK} 	&	\SI{25}{mK} 		&\SI{40}{mK} 		& 	\SI{300}{\mu W}&DR
		\end{tabular}
	\end{center}
	\caption{Temperatures and cooling capacities of the CLASS cryostat stages.  Temperatures given are for the baseplates of each stage, in factory configuration and deployment configuration.  The cooling capacity is the amount of additional power the cooler can accommodate before the temperature rises enough to detrimentally affect the cryo system operation.}
	\label{stages}%
	\vspace{2mm}
	\end{table}
	% CHeck \SI{100}{mK} cooling capacity: ext heater 120 ohm, 3.8V ~ 120 mW

	The Pulse Tube (PT) uses pulses of high-pressure helium (pressurized by a compressor unit) to transport heat out of the system.  The CLASS receivers each use a Cryomech PT 415 paired with a CP1100 series compressor\cite{Cryomech}.  The Pulse Tube cools the \SI{60}{K} and \SI{4}{K} stages.  It is rated for \SI{1.5}{W} of cooling power at the \SI{4}{K} stage, when the stage is at \SI{4.2}{K}, and \SI{40}{W} of cooling power at the \SI{60}{K} stage, when the stage is at \SI{45}{K}.  With no extra load, the PT can cool the \SI{4}{K} stage to \SI{2.8}{K}\cite{Cryomech}.  During typical operation (including the deployment configuration), the CLASS receivers operate with the \SI{4}{K} baseplate at around \SI{3}{K}, and the \SI{60}{K} baseplate at around \SI{40}{K}.

	The Dilution Refrigerator cools the system by mixing $^4$He and $^3$He -- the diluted state has higher enthalpy, so mixing the isotopes takes energy from the surrounding environment.  By pumping on the \emph{still} section of the system, $^3$He is circulated and re-introduced to the mixing chamber, allowing for continuous cooling\cite{bf}.  The DR in the CLASS cryostats cools the \SI{1}{K} stage (where the still is located), and the \SI{100}{mK} stage (where the mixing chamber is located).  The \SI{1}{K} stage operates (including in deployment configuration) at 500-\SI{800}{mK}, and can be electrically heated to encourage $^3$He evaporation and recirculation, increasing the cooling power (to a point).  In factory configuration, the \SI{100}{mK} stage operates at around \SI{25}{mK} and has around \SI{300}{\mu W} of cooling power at \SI{100}{mK}.

\begin{figure}[h!]%
			\begin{center}
			% \begin{subfigure}{.5\textwidth}
				\includegraphics[width=.6\textwidth]{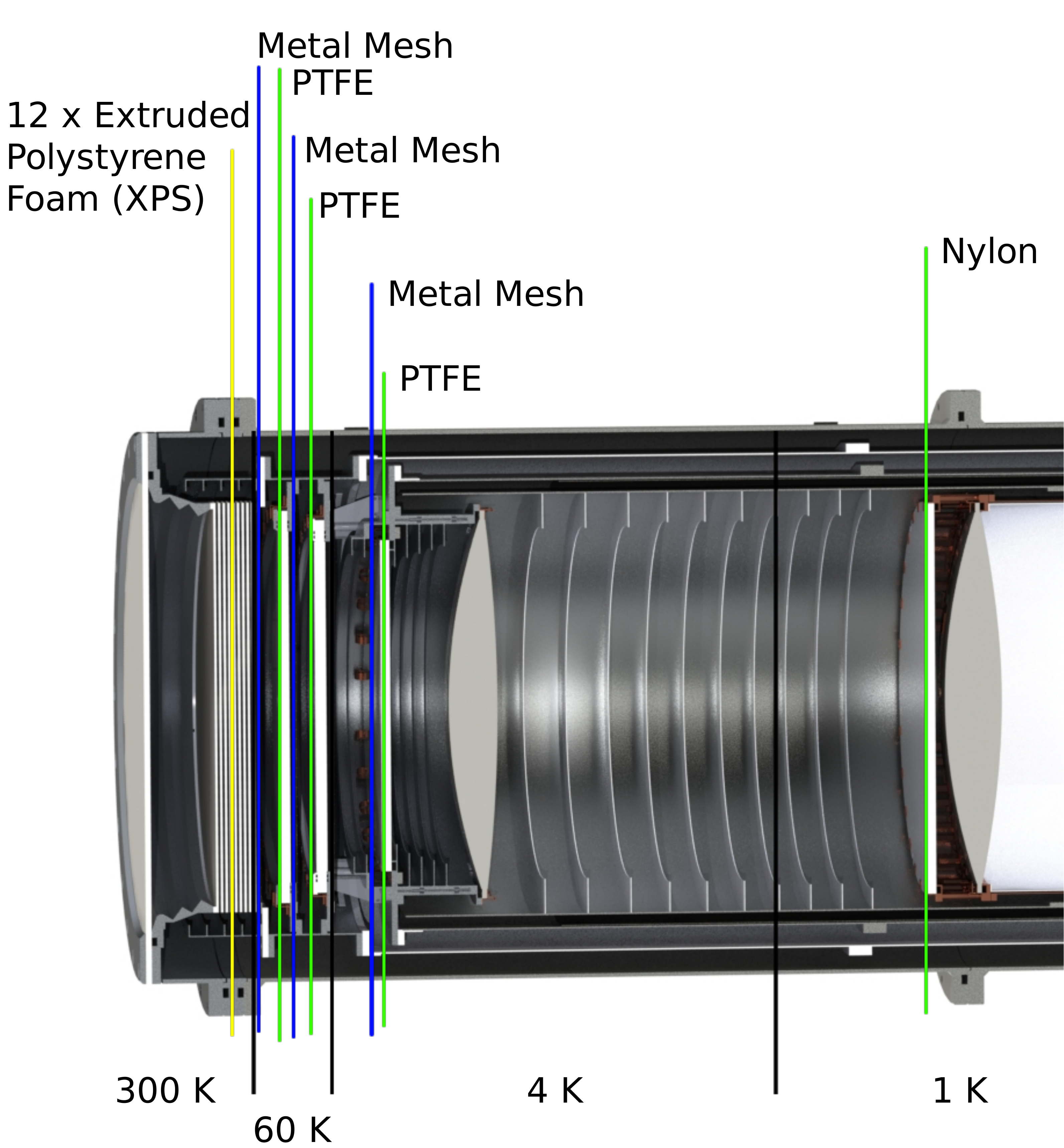}%	
			% \end{subfigure}
			\end{center}
			\vspace{2mm}
			\caption{A diagram of the \SI{90}{GHz} receiver filter and optical design.}%
			\label{wfilterdiagram}
		\end{figure}

\ssection{Optical Design}
	In this section, we will discuss the design, construction, and mounting of the various optical elements (window, lenses, filters) in the CLASS receivers, as well as stray light control.  Figure \ref{wfilterdiagram} shows the \SI{90}{GHz} optical design.   The entire set of filters and lenses should prevent less than 26\% of \SI{90}{GHz}-band light from reaching the detectors (see Table \ref{qfiltertable}).  At the same time, the filters prevent excessive loading at the cold stage.  Deployed on the mount, the \SI{90}{GHz} receiver's \SI{100}{mK} stage is $\sim$\SI{41}{mK}, which would correspond to $<$~\SI{50}{\mu W} of excess loading, mainly from out-of-band infrared radiation (see Figure \ref{cryopower}).  Table \ref{qfiltertable} shows the behavior of the filters and lenses at \SI{40}{GHz} -- the observed \SI{40}{GHz} receiver in-band loss is largely consistent with the design estimate.

		\begin{table}[h]
			% \begin{subfigure}{.7\textwidth}		
			\begin{center}
				\begin{tabular}{llSSSS}
								&Filter Type	& {Temp (K)}		& {Reflectivity (\%)}	& {Absorption (\%)}	& {Spill (\%)}\\
					\SI{300}{K} Stage		& \\
					\hline
					Window 		&				&					& 0.2{-2} 				& 0.23						& 0.025		\\
					XPS (x 12) 	&RT-MLI			& {150 (coldest)} 	& {$<$}0.2 { (each)} 	& 0.3{ (each)} 				& 			\\
								&				& 					& 						&							&			\\
					\SI{60}{K} Stage\\
					\hline
					Metal Mesh 	&Reflective 	&					& 0.5					&							&			\\
					PTFE 		&Absorptive 	& 120 				& 0.2{-1.8} 			& 1.6						& 0.09		\\
					Metal Mesh 	&Reflective 	&					& 0.5					&							&			\\
					PTFE 		&Absorptive 	& 65 				& 0.2{-1.8} 			& 1.6						& 0.081		\\
					\\
					\SI{4}{K} Stage	&\\
					\hline
					Metal Mesh 	&Reflective 	&					& 0.5					& 							&			\\
					PTFE 		&Absorptive 	& 12				& 0.2\text{-1.8} 		& 1.6						& 1.5		\\
					Stop 		&				& 					&						&							& 12.6		\\
					Lens 		&				& 8 				& 0.1 					& 3							& 4.5		\\
					\\
					\SI{1}{K} Stage  	&\\
					\hline
					Nylon 		&Absorptive 	& 1 				& 0.3 					& {$<$} 3 					& 3			\\
					Lens  		& 				& {$<$}1 			& 0.1 					& 4							& 0.3		\\

				\end{tabular}

				% \begin{tabular}{llSS}
				% 		&	\text{Reflectivity (\%)} & \text{Absorption} (\%)\\

				% 	PTFE ? 0.7
				% 	Nylon ? <2

				% 	1K lens ? 1
				% 	4K lens ? 1.5

				% 	mm ? <.5
				% 	XPS  ? too small to measure (<.2)

				% \end{tabular}
			% \end{subfigure}
			\end{center}
			\caption{The \SI{90}{GHz} filters, with the estimated temperature, in-band reflection and absorption, and spill onto non-detector surfaces for the various elements.  For the XPS stack, the reported temperature is of the coldest -- or most interior -- filter.  The reflection and absorption values combine the design and material properties as well as measurements of some components or test pieces.  Total loss to absorption or reflection is less than 26\%.  Including loss due to spill, more than  than 58\% of in-band (and in-beam) light hitting the window reaches the detectors.}%
			\label{wfiltertable}
			
		\end{table}

			\begin{table}[h]
			\begin{center}
				\begin{tabular}{lSS}
									& {Reflectivity (\%)}	& {Absorption (\%)}	\\
					PTFE Filters	& 0.2					& 0.7				\\
					Nylon Filter 	& 0.2 					& {$<$}2			\\
					\SI{1}{K} Lens 		& 0.2					& 1					\\
					\SI{4}{K} Lens 		& 0.2					& 1.5				\\
					Metal Mesh Filters & {$<$}0.5 					& 			\\
					XPS Filters 	& 						&	{$<$}0.2{ (Each)}

			\end{tabular}

			\end{center}
			\caption{Absorption and reflection of filters and lenses at \SI{40}{GHz}.  Including the loss due to spill, the original \SI{40}{GHz} receiver design (before updating it to match the \SI{90}{GHz} filtering scheme) allows an estimated 68\% of in-band light to reach the detectors.  Observation results indicate an achieved \SI{40}{GHz} receiver efficiency of $\sim$60\%\cite{JohnAEfficiency}, which largely matches the design estimate.  The updated \SI{40}{GHz} filter design will slightly increase receiver efficiency.}%
			\label{qfiltertable}
			
		\end{table}

	\ssubsection{\SI{300}{K} Optics}\label{300koptics}
	\label{window}
		The first element in the receiver is the window, which is made of 0.125 inch thick Ultra High Molecular Weight Polyethylene (UHMWPE).  The CLASS windows are Anti Reflection (AR) coated on both sides with a thin, band specific, layer of Polytetrafluoroethylene (PTFE) foam.  This foam is adhered to the surfaces of the window by heating a thin layer of Low Density Polyethylene (LDPE) between the window surface and the PTFE AR coating to act as an adhesive.  Other than frequency-driven differences in the AR coating layer, the window will be nearly the same for all instruments, although the diameter is slightly smaller in \SI{90}{GHz} than the \SI{40}{GHz} design.

	% XPS
	\label{xps}
		In the \SI{90}{GHz} receiver, the next element is a stack of Extruded Polystyrene (XPS) foam filters.  Specifically, we use Model Plane Foam, without coloring or fire retardant which would increase the material's in-band emissivity.  XPS is a light foam, with index close to one for millimeter wavelengths -- hence its reflection in-band is negligible: \SI{90}{GHz} reflection was below what we could measure, bounded at $<0.2\%$.  However, even thin sheets are fairly opaque in the infrared.  So, a stack of XPS sheets has properties very similar to MLI: the radiative power transferred from above the stack to below it is cut by a factor of roughly $1/N$ (where $N$ is the number of filters)\cite{mli,rtmli}.  This XPS filter stack reduces incoming optical power nearly as much as blanking-off the \SI{300}{K} stage with a reflective plate.  Based on this similarity in performance, we can estimate that the coldest (most interior) XPS filter is around \SI{150}{K} ($T_{filter}^4\approx\epsilon (\SI{300}{K})^4$), cutting incoming power to around $5\%$ of what passes through the window.

			\begin{figure*}[h!]%
				\begin{center}
				\includegraphics[width=.8\columnwidth]{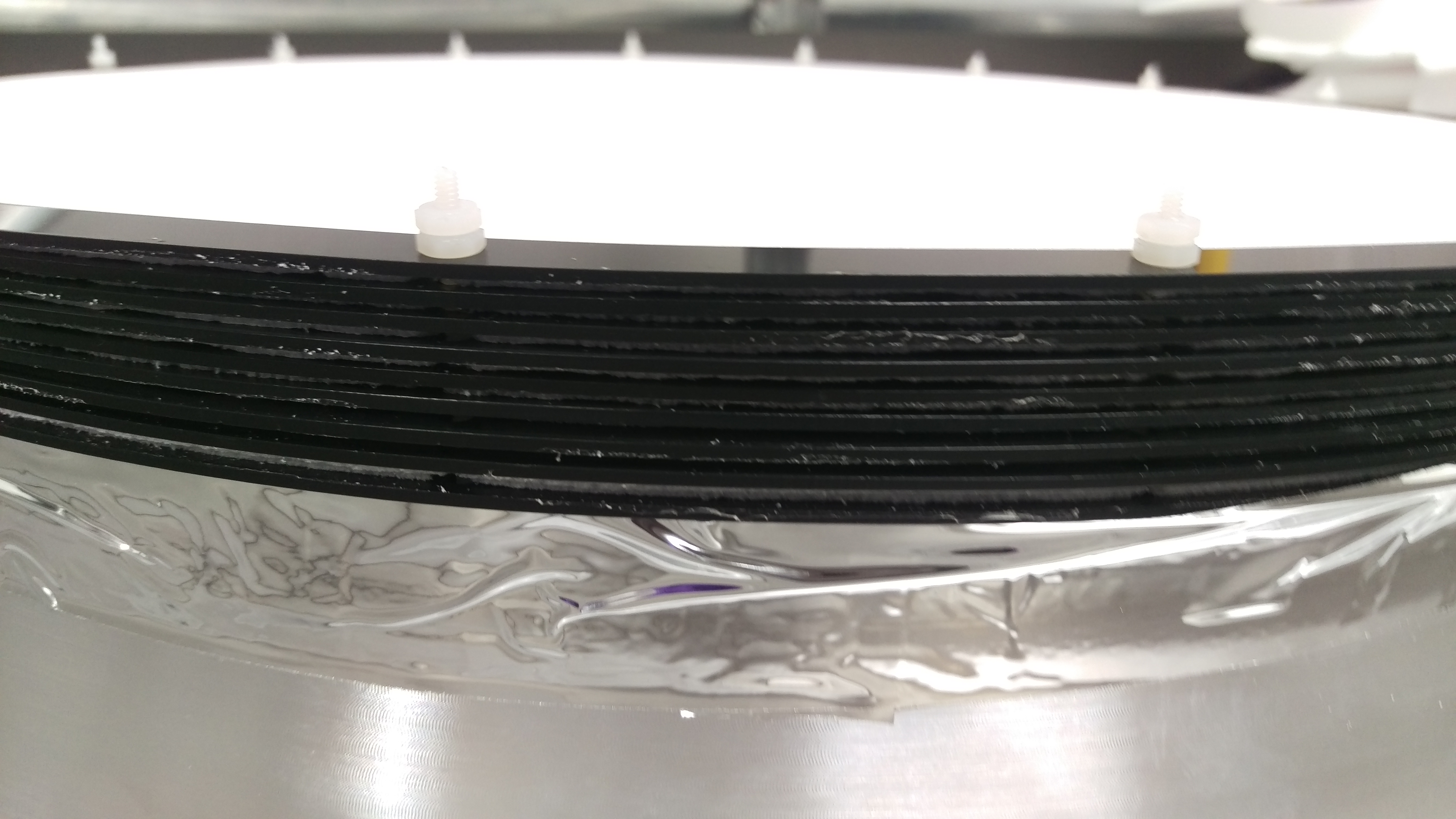}%	
				\end{center}
				\caption{The \SI{300}{K} XPS foam stack for the first \SI{90}{GHz} receiver.  The stack consists of ten $\sim$0.1~in thick sheets of XPS, which were cut to shape on a CNC mill at Johns Hopkins (the same machine used to make the lenses and AR coating\cite{katie}), and two thicker sheets.  The ten thinner sheets are each glued (with rubber cement) to a lasercut Polyoxymethylene (POM) plastic ring.  These rings are then connected to each other, and to the aluminum ring of the \SI{300}{K} stage with threaded nylon rods (and nuts).  Nylon is used because it is transparent and less-reflective in-band than metal.  It is also strong enough to hold the stack together, while also being flexible: the coldest filters will reach $\sim$\SI{150}{K}, meaning the foam will contract by as much as few mm.  The flexibility of the nylon rods allows for differential contraction along the stack.  The plastic rings will contract with the foam, preventing excessive force on the filters or adhesive -- there are also slots rather than clearance holes in the rings to help accommodate contraction.  Two thicker sheets of foam sit in front of the others -- between the stack visible in the picture and the window (which is toward the bottom of the picture).  These are held in place mainly by the geometry of the surrounding ring.}
				\label{xpsstack}%
				\vspace{2mm}
			\end{figure*}

		Unlike absorptive filters, the XPS sheets do not need to be well heat-sunk to the receiver -- in fact, the different sheets are conductively isolated.  The \SI{90}{GHz} XPS stack consists of twelve total sheets (see Figure \ref{xpsstack}).  The \SI{40}{GHz} instrument, when deployed, did not have an XPS stack at \SI{300}{K}.  Given the success of the foam filter stack, during the deployment of the first \SI{90}{GHz} instrument, the \SI{300}{K} filters of the \SI{40}{GHz} instrument were updated to follow the \SI{90}{GHz} design.  The HF instrument will employ a similar strategy for \SI{300}{K} filters. 

	\ssubsection{\SI{60}{K} Optics} \label{60koptics}
	% Metal Mesh
	\label{metal mesh}
		The \SI{60}{K} stack consists of two metal mesh filters and two PTFE filters (see Figure \ref{wfilterdiagram} and Table \ref{wfiltertable}).  The metal mesh filters are largely transparent in-band, and reduce loading by reflecting infrared power.  These filters are described in Section 6.3 of Ref.~\citenum{tom}.  The metal mesh filters are mounted in the receiver by taping\cite{tape} them onto an aluminum ring, putting the filter under enough tension to keep it flat in the ring; the ring is then bolted into the desired location.  We use metal mesh filters from \emph{Tech-Etch}\cite{techetch}, with measured in-band reflection of around $0.5\%$ (depending on the exact design and sample).  The \SI{40}{GHz} receiver's original filter scheme was more conservative (in ensuring filtering infrared power), and included ten metal mesh filters throughout the system as opposed to the \SI{90}{GHz} receiver's three metal mesh filters.  To improve \SI{40}{GHz} optical efficiency, during the deployment of the first \SI{90}{GHz} receiver, many of the metal mesh filters were removed from the \SI{40}{GHz} instrument, migrating it to a filtering scheme that closely mimics the \SI{90}{GHz} approach.

\begin{figure*}[h!]%
				\begin{center}
				\includegraphics[width=.6\columnwidth]{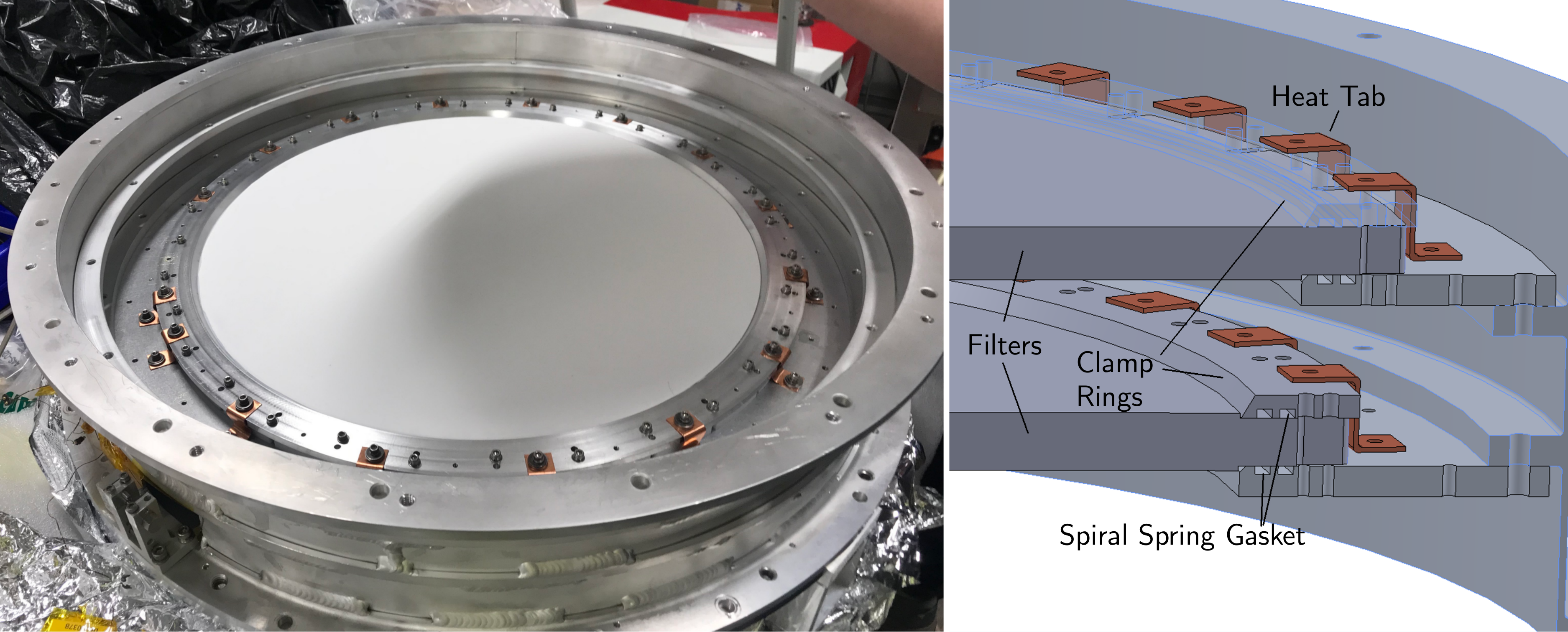}%	
				\end{center}
				\caption{\SI{60}{K} PTFE absorptive filter mounting scheme in the \SI{90}{GHz} receiver.  The filter is sandwiched between two metal discs.  The discs are connected via two concentric bolts circles -- providing an even pressure distribution when the bolts are tightened -- with slots in the filter to accommodate radial thermal contraction.  The interface between each ring and the filter has two Spira\cite{Spira} spiral-shaped ring gaskets, with Apiezon N grease applied\cite{Apiezon}, which fit into the slots visible on the right.  These gaskets allow good thermal contact (even with some vertical -- normal to the filter surface -- contraction), and easily let the filter slide past when it is contracting radially.  To help carry heat out of the filter, the top ring is connected to the rest of the optical stack with copper heat tabs.  The AR coating on the filter ends before the rings, preventing the coating from de-laminating at the edges due to thermal contraction and expansion.}
				\label{60kptfe}%
				\vspace{2mm}
			\end{figure*}

	% PTFE
	\label{PTFE}
		The PTFE absorptive filters are around \SI{9}{mm} thick, and AR coated with an LDPE-adhered layer of PTFE foam (the same material and approach used for the window AR coating described in Section \ref{window}).  PTFE absorbs strongly for frequencies higher than around \SI{1}{THz}, and that power is conducted through the filter, to the metal rings onto which it is mounted, and eventually to the \SI{60}{K} stage of the PT.  In the \SI{40}{GHz} instrument, the \SI{60}{K} PTFE filters were mounted using the same method as the other absorptive filters and the lenses (see Section \ref{1klens} and Figure \ref{lensmount}).  For the \SI{90}{GHz} instruments, the \SI{60}{K} PTFE filters are clamped in place with spiral-spring gaskets (see Figure \ref{60kptfe})\cite{ACT}.  In deployment configuration (but looking at a \SI{300}{K} room rather than the sky), the PTFE filters were estimated to be at \SI{120}{K} and \SI{65}{K} -- but they will be colder during observation due to the reduced radiative loading of the sky and colder environment.
		For the multichroic High-Frequency instrument, we are considering similar PTFE filters or a low-index foam.
	% Arogel
	\label{aerogel}

\begin{figure*}[htb!]%
			\begin{center}
			\includegraphics[width=0.8\columnwidth]{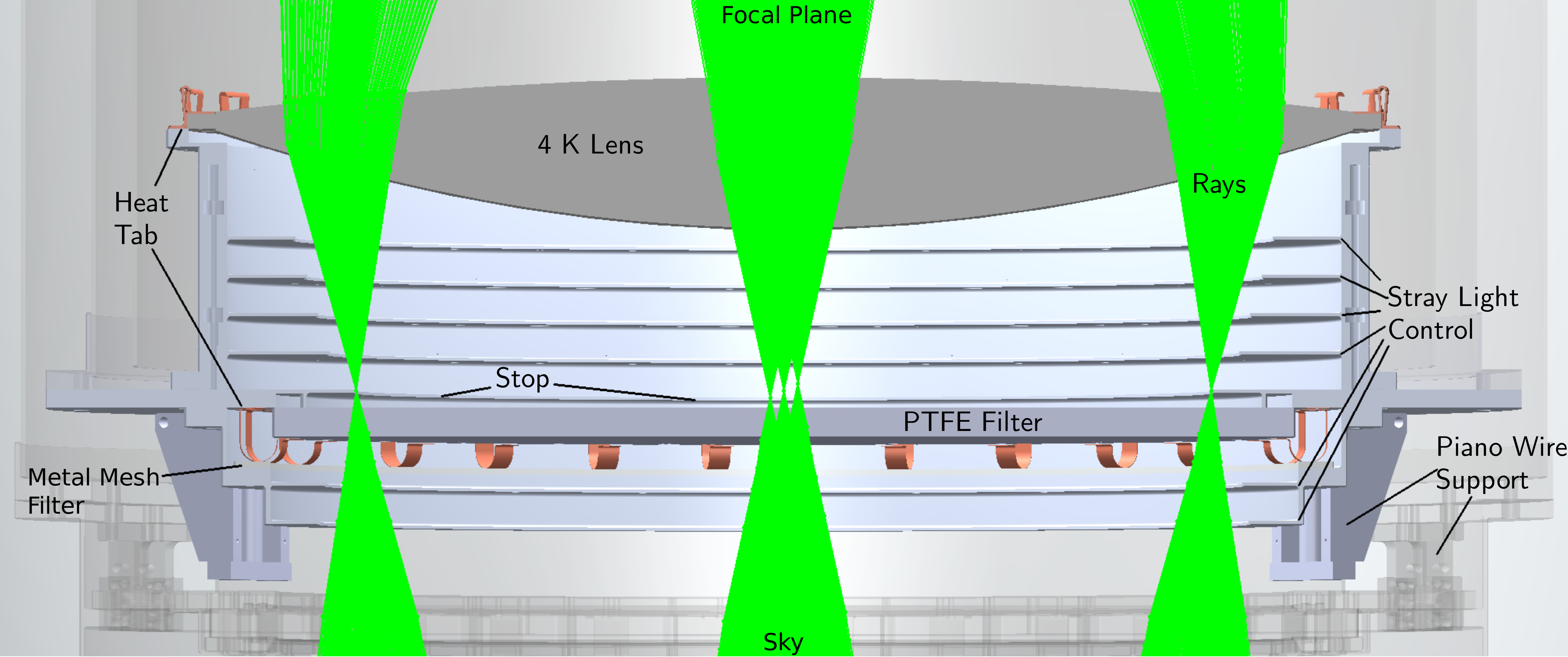}%
			\end{center}
			\caption{The \SI{4}{K} optical stack.  The top is the detector side, and the bottom is the sky side.  A (subset) of rays can be seen in green.  The stack includes: the \SI{4}{K} lens, the stop, the \SI{4}{K} PTFE filter, and a ring for the metal mesh filter.  \SI{60}{K} - \SI{4}{K} piano wire supports are also visible (see Section \ref{PianoWireSupports}).}%
			\label{4krays}%
			\vspace{2mm}
		\end{figure*}

\begin{figure*}[htb!]%
			\begin{center}
			\includegraphics[width=0.8\columnwidth]{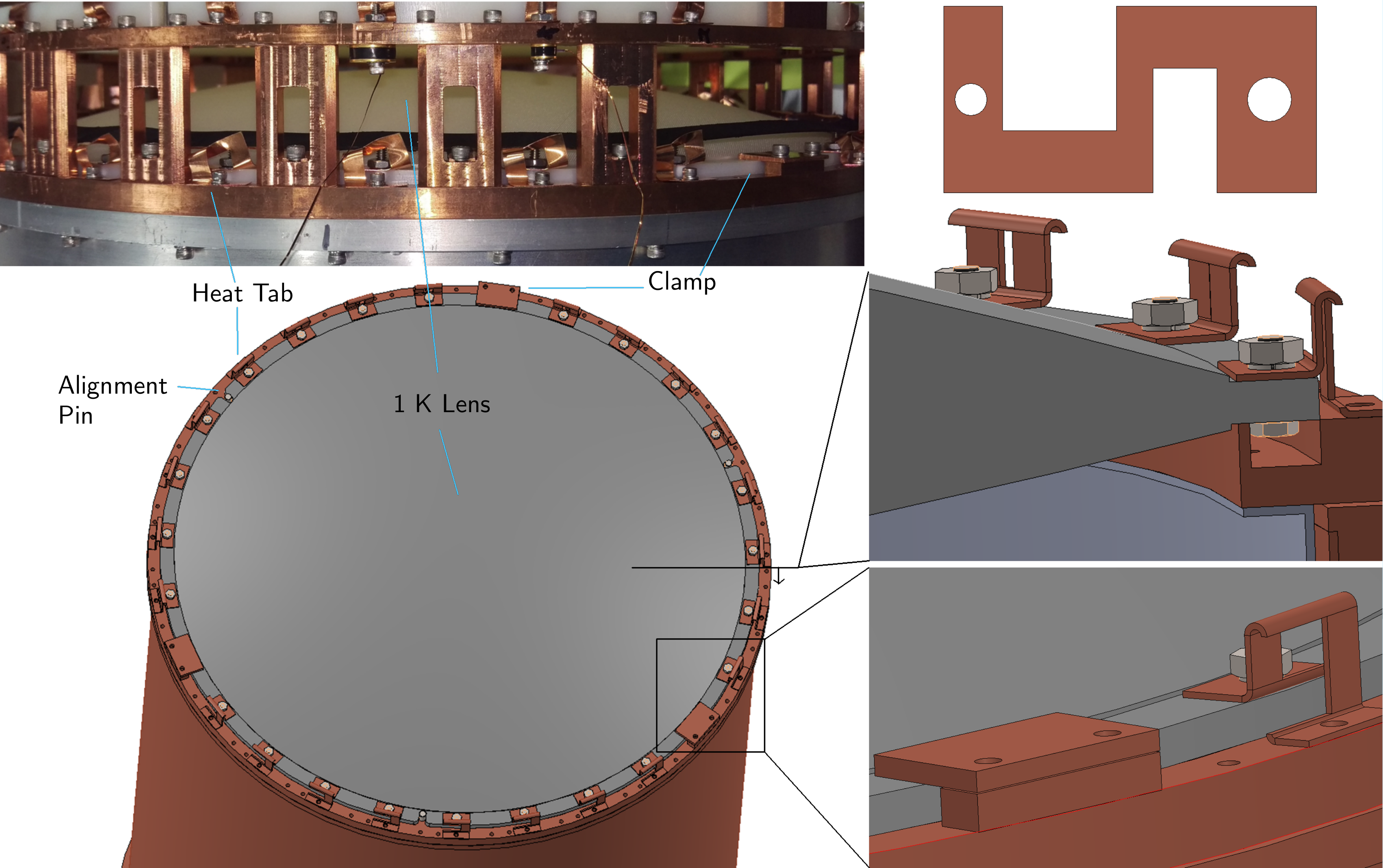}%
			\end{center}
			\caption{Mounting of the \SI{90}{GHz} \SI{1}{K} lens.  The photo on the top left is from the first \SI{90}{GHz} instrument.  Clamps hold the lens onto the \SI{1}{K} ring, while pins that fit into slots on the flange of the lens keep it concentric with the ring.  Copper heat tabs (unbent shape shown in the top right) thermally connect the lens to the ring.  The pins, clamps, and heat tabs accommodate the thermal contraction of the lens.}%
			\label{lensmount}%
			\vspace{2mm}
		\end{figure*}

	\ssubsection{\SI{4}{K} Optics}\label{4koptics}
		\label{lensarcoating}
		The \SI{4}{K} stack has a metal mesh filter (see Section \ref{metal mesh}), PTFE filter (see Section \ref{PTFE}), and the \SI{4}{K} lens -- see Figure \ref{4krays}.  The lens and PTFE filter are mounted in the same way as the \SI{1}{K} lens and filters (see Section \ref{1klens} and Figure \ref{lensmount}).  The CLASS lenses, for \SI{40}{GHz} and \SI{90}{GHz}, are High Density Polyethylene (HDPE), and are AR coated with a simulated dielectric layer.  The simulated dielectric is realized by drilling holes on the surface of the lens (or other element), which are tightly spaced relative to the band wavelength.  This creates a layer with a lower index than the plastic of the optical element, whose properties can be controlled to form an AR coating\cite{tom,katie}.  With AR coating the CLASS lenses should (together) reflect only $0.2\%$ of in-band power.\cite{tom}.  Because the High-Frequency instrument is multichroic -- it has a \SI{150}{GHz} and \SI{220}{GHz} band -- the AR coating has to be multi-layered.  The \SI{4}{K} PTFE filters for the \SI{40}{GHz} and first \SI{90}{GHz} receivers were AR coated with a PTFE foam layer (see Section \ref{window}).  The HF instrument \SI{4}{K} optical stack is similar in design, except that silicon will be used for lens construction rather than HDPE (due to its advantages at the higher frequency band).  These lenses are mounted by clamping them between metal rings with spiral-spring-gaskets, similar to the approach described in Ref.~\citenum{ACT} for mounting silicon lenses, which has been an effective method for holding the \SI{60}{K} PTFE filters in the \SI{90}{GHz} receivers (see Figure \ref{60kptfe}). 

	\ssubsection{\SI{1}{K} Optics}\label{1koptics}
	% 1K lens
	\label{1klens}
		The \SI{1}{K} stage has a nylon absorptive filter and the \SI{1}{K} lens.  The \SI{1}{K} lens is constructed and AR coated just like the \SI{4}{K} lens (see Section \ref{4koptics}).  Figure \ref{lensmount} shows how the \SI{1}{K} lens is mounted (the \SI{4}{K} lens and many of the absorptive filters are mounted similarly).  Concentricity is set by three pins, which fit radial slots on the outside edge of the lenses, allowing for the significant radial thermal contraction while maintaining the center-position of the lens.  The lens is held onto the ring with three clamps.  The clamps are slightly shorter than the lens flange thickness, allowing them to compress the flange and maintain pressure despite thermal contraction in flange thickness.

		Due to the low thermal conductivity of plastics and mass of the lens ($\sim$ \SI{4}{K}g), many oxygen-free high conductivity copper heat tabs are used to thermally connect the lens with the metal ring of the optical stack.  The bolts connecting the heat tabs to the lens are tightened until the plastic of the lens flange is (locally) deformed, leveraging the elasticity of the material to keep good contact when cold.  There is a space cutout for the bolt heads, shaped to accommodate movement as the lens contracts.  Apiezon N Grease\cite{Apiezon} is used to improve thermal conduction between the heat tabs and lens.  The heat tabs are thin (most under \SI{0.01}{in} thick), and shaped for flexibility in both the radial direction and normal to the lens surface.  This prevents the copper tabs from applying stress to the lens as it contracts and expands.  While earlier versions of the tabs were hand-cut from sheets of copper, for the \SI{90}{GHz} receivers most heat tabs are lasercut.  At \SI{1}{K}, the heat tab nuts are covered by a blackened ring to prevent in-band reflections off the more complicated geometry.  The thermal connections provided by the heat tabs allow the lenses to cool in under a week, which is a short time-scale given that the deployed system stays cold for many months at a time. 

	% Nylon
		The \SI{1}{K} nylon filter is mounted in the same way as the lenses (and can be seen at the top of the photo in Figure \ref{lensmount}).  The nylon filter is AR coated with a simulated dielectric layer (like the lenses, see Section \ref{lensarcoating}), with measured reflection of $<$~0.3\% at \SI{90}{GHz} (compared to an expected reflection of $>$~10\% when not AR coated).  Nylon is used as the last filter because it cuts-off closer to the band (reducing potential blue-leak power reaching the detectors, which would increase noise if the detectors couple to it even at a small level).  However, this closer cutoff also leads to increased in-band absorption.  For the HF instrument, nylon is avoided in favor of a low-pass edge filter specifically designed to be compatible with the HF bands.

	\ssubsection{Stray Light Control}

		The interior of the shells and optical stacks -- any surface that is visible to the detectors -- is blackened.  The blackening is a mixture of Stycast\cite{stycast} epoxy with carbon lamp black.  Specifically one ``batch'' would require: \SI{100}{g} 2850~FT, \SI{7.5}{g} 23~LV, \SI{15.78}{g} carbon black.  Rather than use a tool to apply this mixture, we found it most effective to use a gloved hand and fingers.

		There are also various blackened baffling structures that help control stray light.  At the end of the \SI{60}{K} stack, there are a set of baffles whose purpose is primarily to make the surface geometrically black, and extend the \SI{60}{K} cavity around the \SI{300}{K} filters.  On the detector side of the stop (see Figure \ref{4krays}), there are a set of stray-light controlling baffles.  These are each separated by a few wavelengths (so that stray light can enter and be trapped in the cavity), and backed off a few wavelengths from the rays corresponding to the outermost detectors.  In order to symmeterize the effects of these baffles on the rays, baffles are also added on the far side of the stop.  Finally, there are a set of glint-baffles installed in the \SI{4}{K} shell.  The size and spacing of these baffles is geometrically set to prevent single-reflection paths through the space between the lenses\cite{baffle,baffle2}.  Just behind the feedhorn openings of the detectors, there is a field stop to control reflections of light not coupling to the feedhorns.% -- see Figure \ref{fieldstop}.

		% \begin{figure*}[htb!]%
		% 	\begin{center}
		% 	\includegraphics[width=0.4\columnwidth]{"W_Cooldown/field_stop_1"}%
		% 	\hfil
		% 	\includegraphics[height=0.4\columnwidth, angle=90]{"W_Cooldown/field_stop_2"}%
		% 	\end{center}
		% 	\caption{Images of the field stop in the first \SI{90}{GHz} receiver.  The field stop is a blackened copper plate that sits just behind the openings of the detector feed-horns and prevents in-band reflection of light that does not couple to the feed-horns.}%
		% 	\label{fieldstop}%
		% 	\vspace{2mm}
		% \end{figure*}
		% % Maybe get rid of this figure	Add labels to it  Chuck says crop or label?

\ssection{Optical alignment}
	The cryostat and the optics design are such that it is natural to place one lens near the end of the \SI{1}{K} radiation shield, and the second lens near the end of the \SI{4}{K} radiation shield, with the focal plane mounted on the \SI{100}{mK} stage.  This approach allows for a fairly straightforward receiver design -- a single lens and filter assembly is created and added to the end of each shell, and each is (largely) independent.  As can be seen in Figure \ref{diagrams}, each of the shells extends from its corresponding baseplate, which itself is supported by warmer baseplates.  Constraining the relative position of these shells through stiff mechanical supports allows us to align the optical system -- and maintain that alignment through telescope rotations.

	\ssubsection{Tolerances}
	The tolerances for misalignment for the \SI{40}{GHz} and \SI{90}{GHz} instruments are given in Table \ref{tols}.  Given the similarity in design between the \SI{40}{GHz} and \SI{90}{GHz} systems, these tolerances are related directly by the ratio of the wavelengths.  The \SI{40}{GHz} tolerances were determined in Ref.~\citenum{eimer}, via Monte Carlo simulations perturbing all telescope optical elements from prescribed locations and comparing the result to a metric (in this case, a particular Strehl ratio).\cite{eimer_optics,eimer}  The tolerances are not threshold values, but indicate the point past which aspects of image quality could begin to deteriorate. X or Y misalignment could affect the performance of some edge detectors, but would not prevent the survey from proceeding.  Similarly, a small de-focus would not be overly detrimental to large-angular scale pattern recovery.  The High-Frequency receiver is largely similar in design, although the optical design will be faster and the detectors will fill a smaller portion of the focal plane.  A similar tolerance analysis would yield constraints a factor of roughly two smaller than the \SI{90}{GHz} tolerances.

		\begin{table*}[htb]%
			\begin{center}
				\begin{tabular}{l|SS}
				\SI{40}{GHz}& {\SI{1}{K} Lens}& {\SI{4}{K} Lens} \\
				\hline
				X (mm)&        4&5\\
				Y (mm)&        7& 5\\
				Z (mm)&        3& 10\\
				$\phi$ ($^\circ$)&      0.8& 0.4\\
				$\theta$ ($^\circ$)&       1.4& 0.7\\
				\end{tabular}\hfil
				\begin{tabular}{l|SS}
				\SI{90}{GHz} & {\SI{1}{K} Lens}& {\SI{4}{K} Lens}\\
				\hline
				X (mm)&        1.8& 2.2\\
				Y (mm)&        3.1& 2.2\\
				Z (mm)&        1.3& 4.4\\
				$\phi$ ($^\circ$)&      0.36& 0.18\\
				$\theta$ ($^\circ$)&       0.62& 0.31\\
				\end{tabular}

			\end{center}
			\caption{Key optical alignment tolerances for the \SI{40}{GHz} and \SI{90}{GHz} instruments. All values are relative to the focal plane position.  Each lens should be concentric with the focal plane, and offset in Z by a particular distance.  Z is the optical axis of the receiver (specifically, the axis normal to the focal plane), X is (approximately) parallel to the bottom surface of the cryostat cube, and Y is (approximately) vertical.  $\phi$ is the amount the lens is rotated about its own x-axis relative to a plane that is parallel with the focal plane.  Put differently, it is the angle between the lens y-axis and the focal plane Y-axis.  The $\theta$ rotation is about the lens y-axis, or the angle between the lens x-axis and the X-axis.  For reference, the wavelength at \SI{40}{GHz} is \SI{7.5}{mm} and at \SI{90}{GHz} is \SI{3.3}{mm}.  \SI{40}{GHz} tolerances are given in Ref.~\citenum{eimer}.  Due to the similarity between the \SI{40}{GHz} and \SI{90}{GHz} designs, the \SI{90}{GHz} tolerances the \SI{40}{GHz} tolerances scaled by the ratio of the wavelengths.}%
			\label{tols}%
			\vspace{2mm}
		\end{table*}

	\ssubsection{Alignment Design}
		Parts can be machined holding tolerances far tighter than required for the \SI{40}{GHz} or \SI{90}{GHz} optical alignment.  So, rather than employ a design intended to rely on adjustments, the CLASS receivers are designed to align the components solely by assembling the receiver.

		Typical machining operations are able to achieve tolerances of better than 10 thousandths of an inch (.25 mm).  Thus, an assembly that stacked $n$ different parts would hold a tolerance of $.25 \sqrt{n}$~mm, meaning we could stack over 25 parts before reaching even the tightest tolerance in Table \ref{tols}.  Moreover, many of the parts in the optical stacks are thin disks whose thickness should be controllable to much better than 0.25 mm.  Many parts also support (and so set the position) of multiple optical elements, meaning their size does not influence the relative position of those elements.

		The fabrication of the shells and some of the cylindrical optical stacks involve welding, which can be detrimental to precision.  The smaller parts -- like the \SI{60}{K} optical stack -- can be milled to the proper height and parallelism constraints after welding.  The shells, however, are far too big for this process to be plausible.  Instead, we use a long inside-micrometer to measure the actual shell length and parallelism before we finalize the design and order the optical stack parts.

		The majority of the components in the receiver are connected to each other by bolts.  While clearance holes can create some positioning uncertainty (and non-repeatability), a bolt pattern with many holes can highly constrain the position of parts (relative to alignment constraints).

	\ssubsection{Measured Alignment} \label{Measured Alignment}
		If the optical stacks are designed, built, and assembled correctly, the receiver optics \emph{should} be aligned.  To verify alignment, we used a coordinate measuring machine -- specifically a FaroArm \cite{faro}.  The receivers are shipped to the field partly disassembled (each shell and the DR are removed to keep the components safe during shipping), and are reassembled at the site -- without access to the FaroArm measuring tool.  So, the receiver was twice assembled and measured before deployment to check that variation with each assembly would be small relative to the tolerances.  Table \ref{Alignment} shows the measurement results compared to tolerances for the first \SI{90}{GHz} receiver.  When the system is fully assembled, the optical elements are inaccessible, and the addition of each shell and optical assembly can bend the supports connecting the baseplates.  This effect is accounted for in the alignment measurements, but the difficulty in measuring it dominated the variation between measurements (meaning that the reported variation does not likely represent actual variation in position).  The measurements reflect the expected cold position of elements, accounting for thermal contraction.  The measurement results demonstrate acceptable alignment (note that the tolerances are characteristic rather than threshold values).

		\begin{table*}[htb]%
			\begin{center}
				\begin{tabular}{l|cr}
				\SI{1}{K} Lens& Tolerance& Measured \\
				\hline
				X (mm)	&  1.80 &	 -0.268 $\pm$  0.390 \\
				Y (mm)	&  3.10 &	 -0.026 $\pm$  1.791 \\
				Z (mm)	&  0.90 &	  0.758 $\pm$  0.698 \\
				$\phi$ ($^\circ$)	&  0.62 &	 -0.132 $\pm$  0.109 \\
				$\theta$ ($^\circ$)	&  0.36 &	 -0.044 $\pm$  0.064 \\
				\end{tabular}\hfil
				\begin{tabular}{l|cr}
				\SI{4}{K} Lens& Tolerance& Measured \\
				\hline
				X (mm)	&  2.20 &	 -1.073 $\pm$  0.509 \\
				Y (mm)	&  2.20 &	 -2.305 $\pm$  1.317 \\
				Z (mm)	&  4.40 &	  4.155 $\pm$  0.682 \\
				$\phi$ ($^\circ$)	&  0.31 &	 -0.278 $\pm$  0.191 \\
				$\theta$ ($^\circ$)	&  0.18 &	 -0.067 $\pm$  0.114 
				\end{tabular}
			\end{center}
			\caption{Expected cold position of the main optical elements for the first \SI{90}{GHz} receiver (based on FaroArm\cite{faro} measurements).  All values are relative to the (measured) focal plane position and orientation (the focal plane would have X, Y, Z, $\phi$, $\theta$ = 0).  Tolerances are from Table \ref{tols}, and the coordinate system is the same as described there.  The values and uncertainty come from combining two sets of measurements -- each taken during a different receiver assembly to ensure consistency in the assembly process.  Each value is the mean of the result from the two measurements, and the reported uncertainty is half the difference between the two results.  The observed variation is likely dominated by measurement uncertainty rather than actual changes in alignment between assemblies.}%
			\label{Alignment}%
			\vspace{2mm}
		\end{table*}

		\begin{figure*}[ht]%
			\begin{center}
			\includegraphics[width=.4\columnwidth]{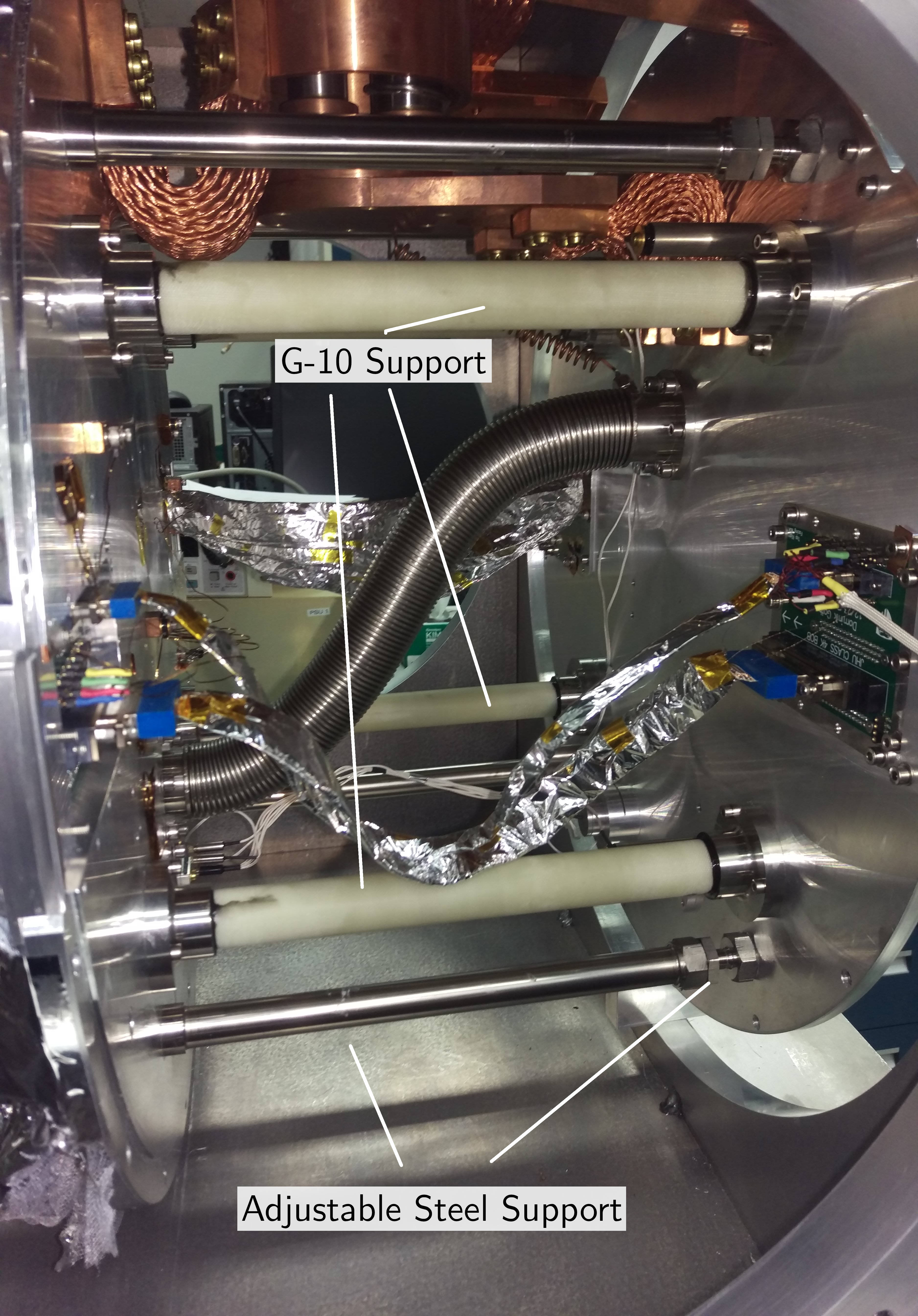}%	
			\end{center}
			\caption{Image of G-10 and adjustable supports in the first \SI{90}{GHz} receiver.  The G-10 supports add significant stiffness -- reducing the movement of the \SI{4}{K} baseplate due to the weight of the shell and optics by a factor of five -- but they add only \SI{0.25}{W} of conductive power to the \SI{4}{K} stage. }%
			\label{g10}%
			\vspace{2mm}
		\end{figure*}

	\ssubsection{Boresight Rotation and Mechanical Supports}
	  
		The CLASS telescopes can rotate in boresight, a feature necessary to measure both Stoke's Q and U and useful for understanding and removing systematics.  However, it also rotates the receiver relative to gravity (see instrument schematic from Figure 5 in Ref.~\citenum{tom}).  This rotation adds a further challenge for optical alignment: components need not only be aligned, but must also stay aligned when the telescope is in different orientations.  This means that the amount by which supports bend due to the load placed on them must be smaller than the alignment tolerances.

		Thin-walled stainless steel tubes make up the majority of mechanical supports in the system -- including the heat switches between the colder stages.  Between the \SI{60}{K} and \SI{4}{K} baseplates, the factory configuration steel supports allowed the \SI{4}{K} baseplate to drop by as much as \SI{5}{mm} when the \SI{4}{K} shell and optics were attached.  These supports were replaced with a new set: four G-10 supports and four adjustable-length stainless steel supports (except in the \SI{40}{GHz} receiver, where the four G-10 supports were simply added to the existing steel supports).

		The G-10 supports can be seen in Figure \ref{g10}.  Each G-10 rod is 1~inch in diameter and 0.5~inches thick, conducting only a collective \SI{0.25}{W} of power from \SI{60}{K} to \SI{4}{K} (G-10 conductivity from Ref.~\citenum{NIST}).  The rods were cut to length on a lathe (at very low speed).  The distance between the \SI{60}{K} and \SI{4}{K} baseplates does not set an important distance in optical alignment: the position of the \SI{60}{K} stack filters relative to the rest of the optics is not highly sensitive to z-direction shifts.  So, while the rods should all be the \emph{same} length, that length does not need to be exact.  Each rod is (after roughening the surface) epoxied into a stainless steel holder on each end using Styacst 2850 FT\cite{stycast}.  While the epoxy remains flexible, the rods are installed between the baseplates, allowing the holders to rotate independently for alignment with bolt holes.  Then, the adjustable-length steel supports are shortened until there is a compressive load on the G-10 rods.  The epoxy is allowed to cure in this final configuration.

		 \label{bending}
		This new set of supports significantly improves the rigidity of the full support structure, reducing the sag of the \SI{4}{K} baseplate to around \SI{1}{mm}.  Table \ref{drop} shows the estimated displacement of the baseplates and optical elements due to the weight of the \SI{4}{K} (and \SI{1}{K}) shell and optical stack, based on FaroArm\cite{faro} measurements.  This displacement corresponds to the amount these elements could shift due to boresight rotation of the telescope (without the piano wire supports -- see Section \ref{PianoWireSupports}).  Despite fairly large measurement uncertainty, these displacements are mostly consistent with alignment tolerances.

		\begin{table*}[!ht]%
			\begin{center}
				\begin{tabular}{c|r}
				\SI{1}{K} Baseplate&\multicolumn{1}{c}{Change}\\
				\hline
				$\Delta$X (mm)	&  0.246 $\pm$  0.158 \\
				$\Delta$Y (mm)	& -1.305 $\pm$  0.673 \\
				$\Delta$Z (mm)	&  0.064 $\pm$  0.044 \\
				$\Delta$$\phi$ ($^\circ$)	&  0.084 $\pm$  0.206 \\
				$\Delta$$\theta$ ($^\circ$)	&  0.099 $\pm$  0.115 
				\end{tabular}\hfil
				\begin{tabular}{c|r}
				\SI{4}{K} Baseplate& \multicolumn{1}{c}{Change} \\
				\hline
				$\Delta$X (mm)	&  0.033 $\pm$  0.166 \\
				$\Delta$Y (mm)	& -1.020 $\pm$  0.009 \\
				$\Delta$Z (mm)	&  0.000 $\pm$  0.000 \\
				$\Delta$$\phi$ ($^\circ$)	&  0.103 $\pm$  0.215 \\
				$\Delta$$\theta$ ($^\circ$)	&  0.064 $\pm$  0.092 
				\end{tabular}\hfil
				\begin{tabular}{c|r}
				\SI{60}{K} Baseplate& \multicolumn{1}{c}{Change} \\
				\hline
				$\Delta$X (mm)	&  0.033 \\
				$\Delta$Y (mm)	& -1.020 \\
				$\Delta$Z (mm)	&  0.000 \\
				$\Delta$$\phi$ ($^\circ$)	&  0.103 \\
				$\Delta$$\theta$ ($^\circ$)	&  0.064 
				\end{tabular}

				\vspace{3mm}

				\begin{tabular}{c|r}
				Effect on Focal Plane& \multicolumn{1}{c}{Change} \\
				\hline
				$\Delta$X (mm)	&  0.794 $\pm$  0.792 \\
				$\Delta$Y (mm)	& -0.839 $\pm$  1.810 \\
				$\Delta$Z (mm)	&  0.062 $\pm$  0.044 \\
				$\Delta$$\phi$ ($^\circ$)	&  0.084 $\pm$  0.206 \\
				$\Delta$$\theta$ ($^\circ$)	&  0.099 $\pm$  0.115 
				\end{tabular}\hfil
				\begin{tabular}{c|r}
				Effect on \SI{1}{K} Lens& \multicolumn{1}{c}{Change} \\
				\hline
				$\Delta$X (mm)	&  0.899 $\pm$  1.080 \\
				$\Delta$Y (mm)	&  0.361 $\pm$  2.913 \\
				$\Delta$Z (mm)	& -0.007 $\pm$  0.000 \\
				$\Delta$$\phi$ ($^\circ$)	&  0.103 $\pm$  0.215 \\
				$\Delta$$\theta$ ($^\circ$)	&  0.064 $\pm$  0.092 
				\end{tabular}

			\end{center}
			\caption{The weight of the shells and optical stacks in the fully assembled receiver can change the position and orientation of structures relative to when the receiver is open (and they are accessible).  The estimated size of this effect -- which also bounds the shifting in optical alignment due to boresight rotation -- is reported here based on FaroArm\cite{faro} measurements.  The top tables show the change in position and orientation of the three baseplates to which shells attach.  The bottom tables show the effect this has on the position and orientation of the focal plane and \SI{1}{K} lens.  The \SI{4}{K} lens is not included as the movement of the \SI{60}{K} plate was minimal (and is in any case irrelevant to the relative alignment of the lenses and focal plane).  The same coordinate system is used as in Table \ref{tols} and \ref{Alignment}.}%
			\label{drop}%
			\vspace{2mm}
		\end{table*}
		% Add image of:A diagram of the parameters you measure, cryostat with some axis and component would make it easier to follow. Imagine someone else has to present this paper at morning coffee, looking at the tables is not useful, but one nice diagram would allow for much more discussion

	\ssubsection{Piano Wire Supports}\label{PianoWireSupports}
		While the boresight dependent shifts in alignment (see Section \ref{bending} and Table \ref{drop}) are not large enough to put the system out of alignment, changes in relative optical element position with boresight rotations remain undesirable.  This shifting of the various elements is due mainly to the cantilevered structure of the system, so additional support at the end of each shell was deemed beneficial.

		\begin{figure*}[h!]%
			\begin{center}
				\includegraphics[width=.9\textwidth]{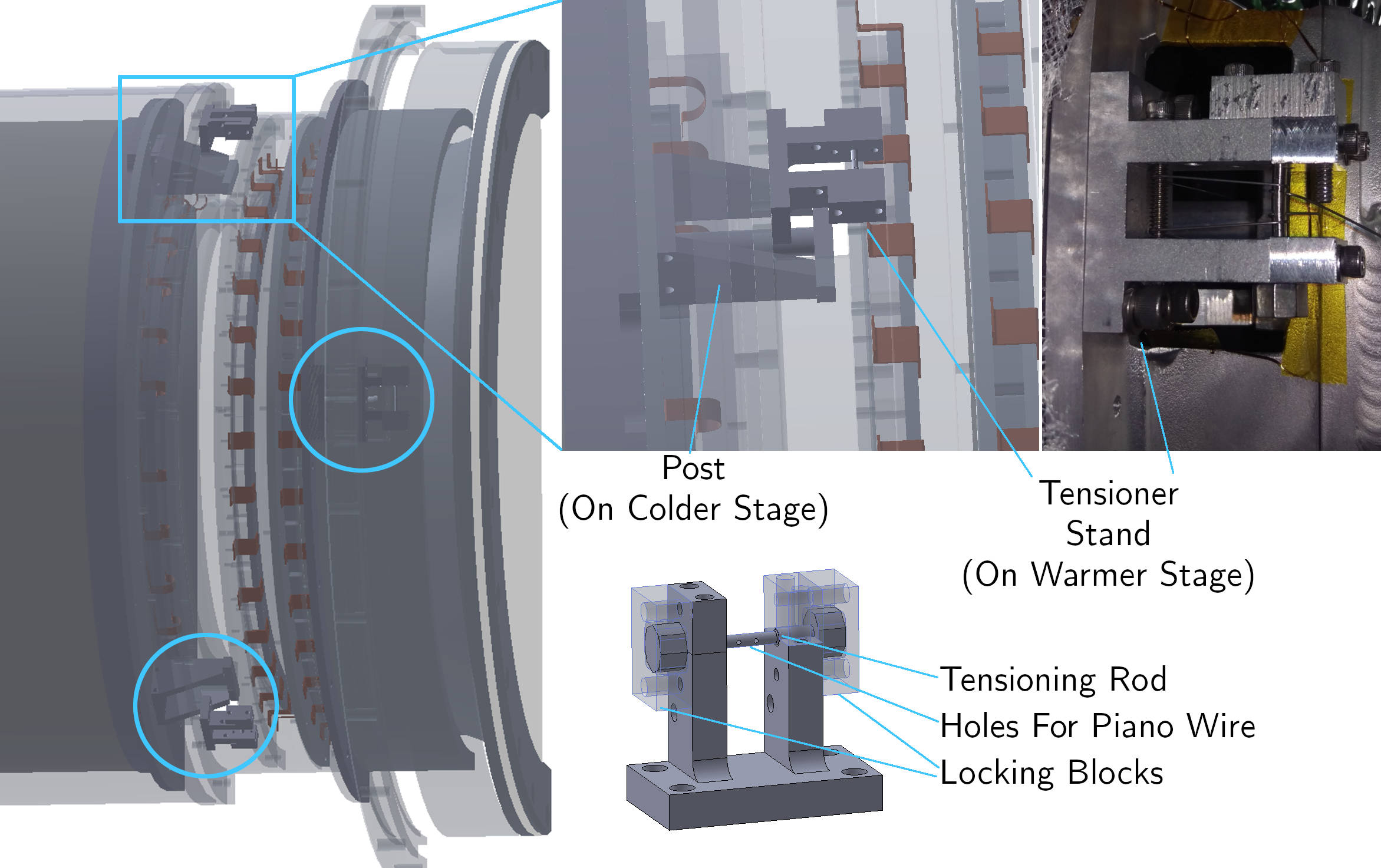}%
			\end{center}		
			\caption{Piano wire supports provide a mechanical connection between the \SI{300}{K} and \SI{60}{K} stages, and the \SI{60}{K} and \SI{4}{K} stages, at the ends of the shells.  Tempered high carbon steel piano wire is run around a post on the colder stage, to a tensioner on the warmer stage.  The tensioner has a steel rod; the wires are run through holes in the rod and then around it as it turns.  When desired tension is achieved, the rod is locked into place with locking blocks that fit around the octagonal ends of the tensioning rod.  and through holes on a rotating tensioning rod on the warmer stage. The photo on the right shows the 60-\SI{4}{K} support tensioned during deployment assembly of the first \SI{90}{GHz} receiver.}%
			\label{pianowire}%
			% \vspace{2mm}
		\end{figure*}

		To that end, we added six piano wire supports to the system (see Figure \ref{pianowire}):  three equally spaced connecting the \SI{300}{K} stage to the \SI{60}{K} stage, and three connecting the \SI{60}{K} stage to the \SI{4}{K} stage.  The wire is \SI{0.6}{mm} in diameter tempered high-carbon (``piano wire'') steel, and the length of wire between the tensioner-rod and post is around \SI{2}{cm} for the 300-\SI{60}{K} supports, and \SI{4.5}{cm} for the 60-\SI{4}{K} supports.  The piano wire supports should (in total) add less than \SI{2}{W} of power to the \SI{60}{K} stage, and less than \SI{0.06}{W} at the \SI{4}{K} stage\cite{NIST}, which is supported by their observed minimal impact on cryogenic performance.

		When properly tensioned, the supports do not change the position of any elements in the receiver (they are not used, for example, to attempt adjustments for alignment).  However, they greatly reduce the potential for movement of elements when the system rotates in boresight.  They are also easy to remove and re-install when opening or closing the system.

\ssection{Magnetic Shielding}

	The CLASS detectors are superconducting Transition Edge Sensor (TES) polarization sensitive bolometers which are inductively coupled to a  multi-stage Superconducting Quantum Interference Device (SQUID) for readout\cite{John_SPIE,Sumit,aamir}. The TES superconducting transition is shifted by coupling to external magnetic fields\cite{aamir}, an effect which changes the responsivity of the detector, mimicking a spurious signal. Moreover, external magnetic field coupling to the SQUIDs almost exactly mimics the inductive coupling from the TES bolometers. In combination, these effects can lead to a pernicious scan-synchronous pickup as telescope azimuth scans swing through the Earth's geomagnetic field, with an ambient value of 23 $\mu$T\cite{mag_field}. 

		\begin{figure*}[htb]%
		\begin{center}
		\includegraphics[width=.8\columnwidth]{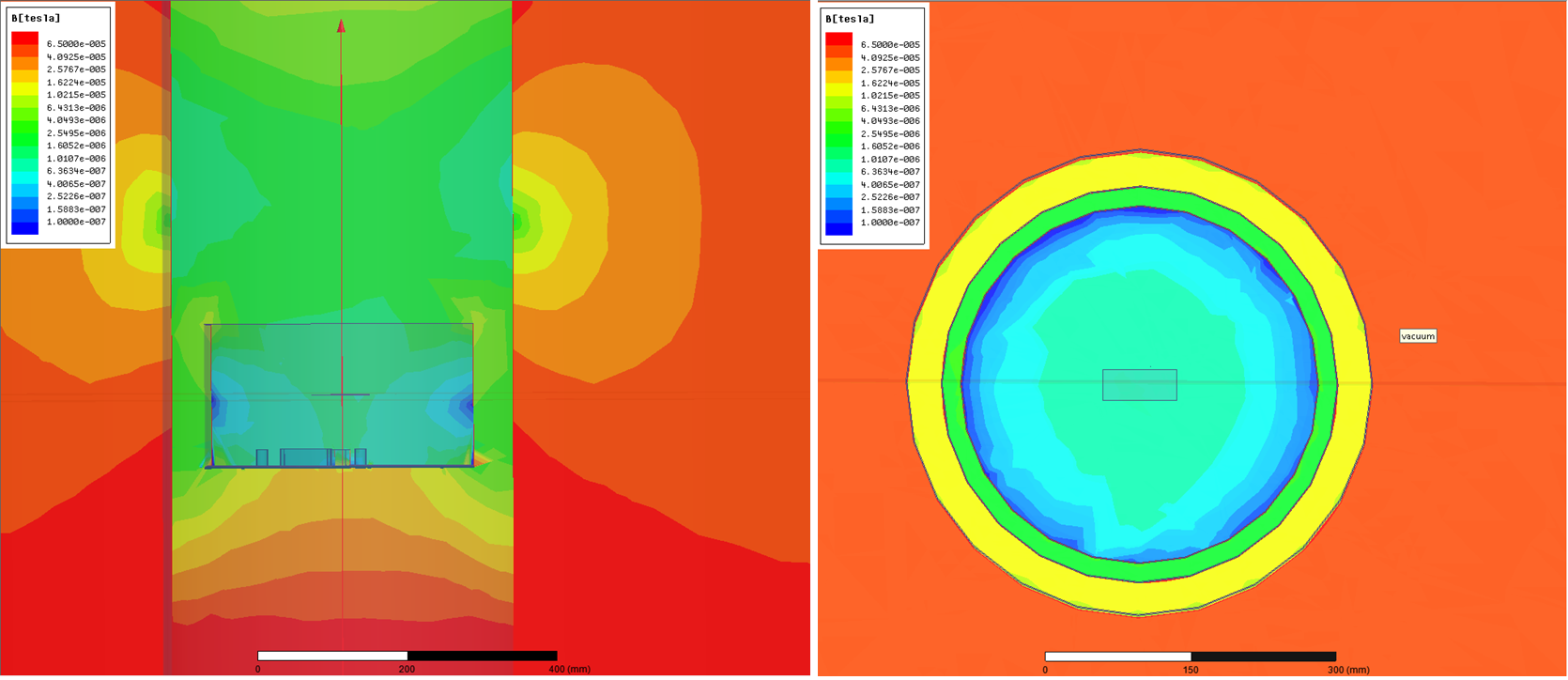}%
		\end{center}
		\caption{Results of finite element analysis with Ansys MAXWELL\cite{Ansys} of \SI{90}{GHz} receiver magnetic shielding design based on \SI{4}{K}, \SI{1}{K}, and cold-stage shell and plate shields (excluding niobium sheeting). The modeled ambient magnetic field is pointed along the axis of the shielding, which is the direction of greatest susceptibility of the magnetic shielding. Near the location of the focal plane (indicated by the rectangle) the ambient magnetic field is attenuated by no less than a factor of 100.}%
		\label{mag}%
		\vspace{2mm}
		\end{figure*}

	Of the two effects, the more sensitive pickup comes through the SQUID readout, which will see any magnetic field greater than 200 nT\cite{Hollister}.  Thus, the receiver design includes magnetic shielding to suppress ambient magnetic fields by at least a factor of 100.  This shielding consists of cylinders constructed of Amumetal 4K high-permeability $\mu$-metal annealed for high performance at cryogenic temperatures\cite{Amuneal}. One cylindrical shield is co-mounted with the \SI{4}{K} radiation shield, providing a large volume of magnetic field attenuation.  A second shield is mounted close to and around the entire focal plane assembly.  The geometry of the \SI{90}{GHz} focal plane is less favorable to magnetic shielding, so in the \SI{90}{GHz} receivers, a third shield is integrated into the \SI{1}{K} shell.

	The magnetic shielding strategy was designed and evaluated with detailed simulations -- performed with the ANSYS MAXWELL software suite\cite{Ansys} -- which show the design is consistent with target suppression (see Figure \ref{mag}).  Limited in-lab testing also supports the effectiveness of the shielding\cite{aamir}.  In addition, the CLASS focal plane modules incorporate an additional layer of 0.\SI{5}{mm} of Niobium superconducting shielding above and below the SQUID readout circuit to further suppress magnetic fields\cite{aamir,Sumit}.

		\begin{figure*}[!htb]%
		\begin{center}
		\includegraphics[width=.5\columnwidth]{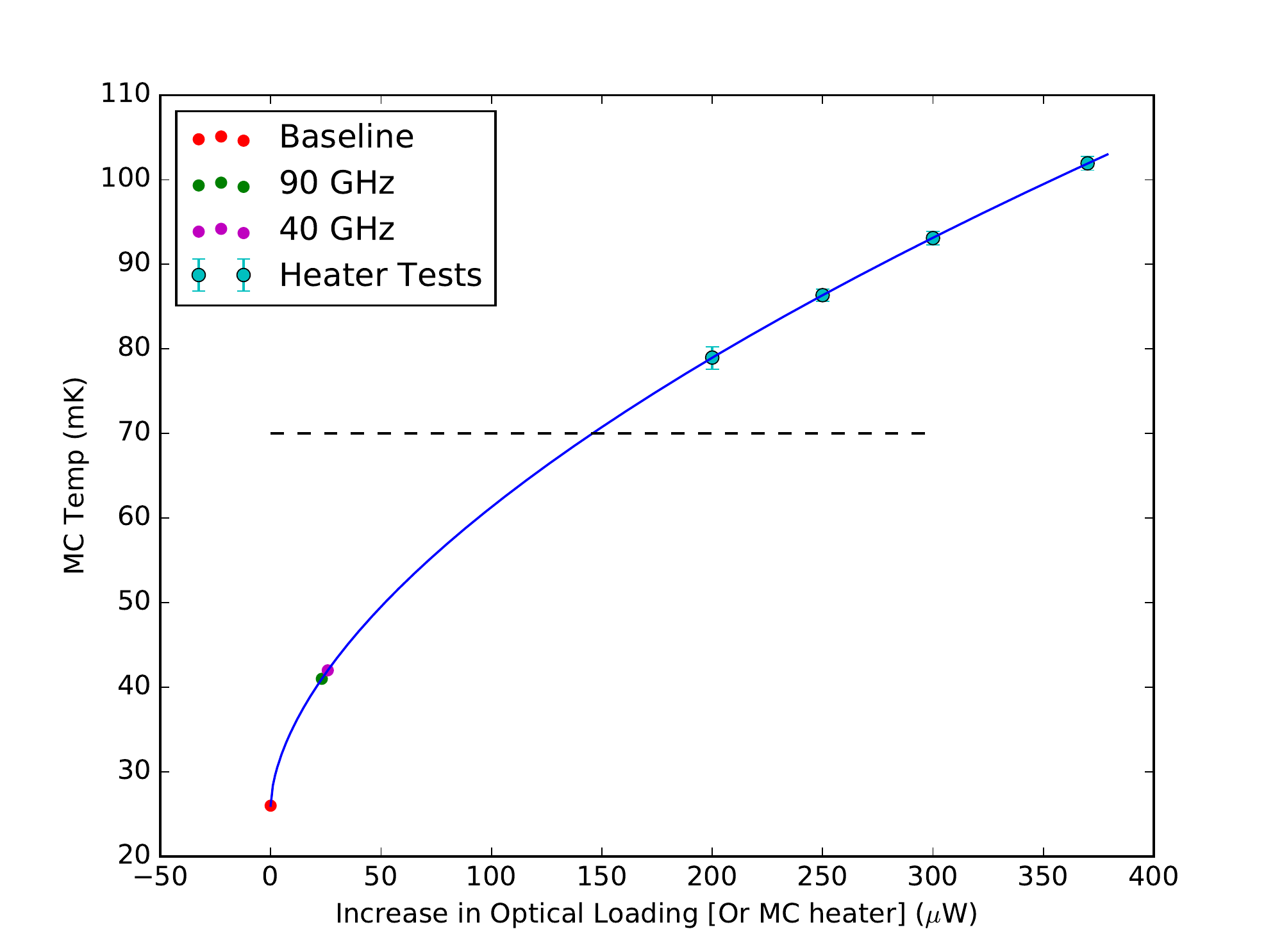}%
						% \hfil
		\includegraphics[width=.5\textwidth]{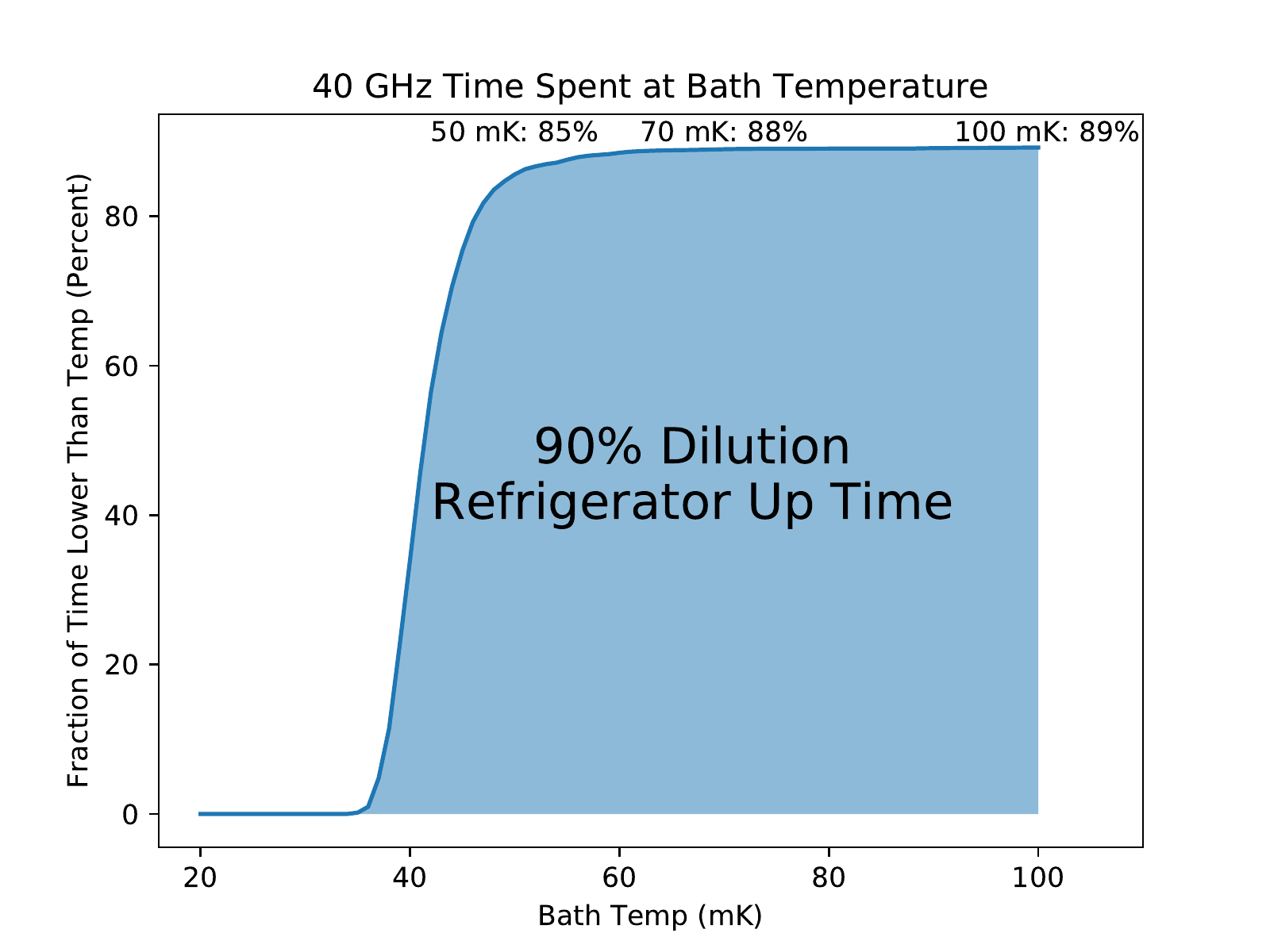}%
		\end{center}
		\caption{{\bf Left:} The temperature of the Mixing Chamber (MC), or \SI{100}{mK} stage of a CLASS cryostat compared to the excess loading (optical or, for testing, heat from a resistor).  The red point is the baseline performance of the system: with zero excess loading, the system can cool to \SI{26}{mK}.  The ``heater test'' points show the results of electrically heating the \SI{100}{mK} stage with a known amount of power from a resistor (with the system in nearly factory configuration).  From this information, we estimate the temperature associated with intermediate levels of loading.  The blue curve is a simple power-law fit to the points.  The \SI{90}{GHz} and \SI{40}{GHz} had average cold stage temperatures when looking at the sky of around \SI{41}{mK} and \SI{42}{mK} respectively.  Thus, we estimate the cold stage loading is \SI{25}{\mu W}, compared to $\sim$1\SI{50}{\mu W} allowable before the temperature rises to \SI{70}{mK}.  {\bf Right:} The total fraction of the time since \SI{40}{GHz} observing began where the cold-stage is below each temperature.  This demonstrates the consistent performance of the system.}%
		\label{cryopower}%
		\vspace{2mm}
		\end{figure*}

\ssection{Field Performance}

\begin{figure*}[!h]%
			\begin{center}
				\includegraphics[width=.9\textwidth]{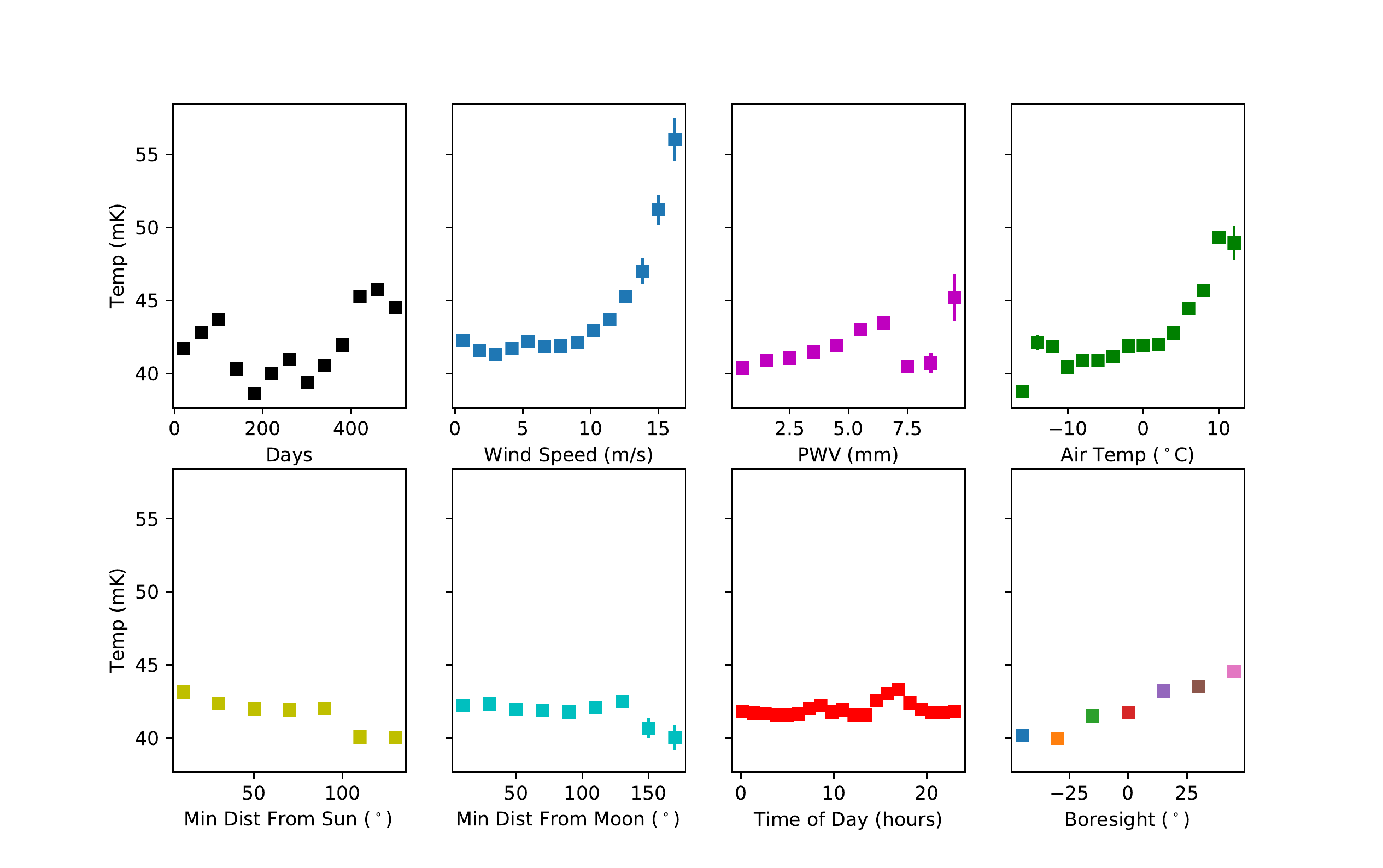}%
			\end{center}		
			\caption{Temperature of the \SI{40}{GHz} cold stage compared to various factors that might affect the system.  Data is taken from the majority of the time the instrument has been in the field, but is limited only to times when it is observing.  Some outlier events have been removed, for example when the cooling system was turned off or lost power during a scan.  The top left shows the temperatures over time, measured in days since the first data point.  Distances from the sun or moon are the minimum angular distance from a single scan in azimuth (temperature data are average values from individual azimuth sweeps).  PWV is Precipitable Water Vapor in the atmosphere.  Time of day is local.  There are also correlations between many of these factors.}%
			\label{qfocaltemps}%
			% \vspace{2mm}
		\end{figure*}

	After over two years of the \SI{40}{GHz} instrument operating in the field, we examine the cryogenic performance while observing for a sustained period.  Figure \ref{qfocaltemps} summarizes the \SI{40}{GHz} cryogenic performance, showing the cold stage temperature with respect to various factors that influence loading on the system.  The cold stage has an average operating temperature of \SI{42}{mK} during observations, and the DR has been operating for nearly 90\% of the time since the instrument began observing.  Temperatures after \SI{40}{GHz} filters were updated are consistent with this performance.  Since deployment, the first \SI{90}{GHz} instrument has operated at around \SI{41}{mK}.  Observations -- particularly scans of the moon -- can be used to understand the instrument beam and pointing.  The results for the \SI{40}{GHz} instrument demonstrate successful and robust optical alignment in the receiver\cite{moonscans}. 

\newpage
\ssection{Conclusion}
	The receiver design successfully addressed the needs of CLASS.  The cryogenic performance is robust, with the \SI{40}{GHz} receiver consistently operating well below the target cold-stage temperature of \SI{70}{mK}, and the first \SI{90}{GHz} receiver demonstrating similar performance since its deployment.  The optical system is high-efficiency, losing less than 26\% of in-band power to reflection or absorption in filters and cold refractive optics at \SI{90}{GHz}.  The key optical elements are aligned within design tolerances and stay well-aligned during telescope rotations in boresight.  The receiver employs stray-light controls, and shields detectors and SQUIDS from the effects of Earth's magnetic field.  For each of CLASS's four telescopes, the receiver -- implemented with this design -- is supporting high-sensitivity observation that enables CLASS to successfully pursue its science goals.

\newpage

\acknowledgments % equivalent to \section*{ACKNOWLEDGMENTS}       
 We acknowledge the National Science Foundation Division of Astronomical Sciences for their support of CLASS under Grant Numbers 0959349, 1429236, 1636634, and 1654494. The CLASS project employs detector technology developed under several previous and ongoing NASA grants. Detector development work at JHU was funded by NASA grant number NNX14AB76A. K. Harrington is supported by NASA Space Technology Research Fellowship grant number NX14AM49H. T. Essinger-Hileman was supported by an NSF Astronomy and Astrophysics Postdoctoral Fellowship. We further acknowledge the very generous support of Jim and Heather Murren (JHU A\&S 88), Matthew Polk (JHU A\&S Physics BS 71), David Nicholson, and Michael Bloomberg (JHU Engineering 64). CLASS is located in the Parque Astron\'omica Atacama in northern Chile under the auspices of the Comisi\'on Nacional de Investigaci\'on Cient\'ifica y Tecnol\'ogica de Chile (CONICYT).

\newpage
% References
\bibliography{spie2018} % bibliography data in report.bib
\bibliographystyle{spiebib} % makes bibtex use spiebib.bst

\end{document}